\pdfoutput=1

\documentclass[11pt]{article}

\usepackage[preprint]{acl}
\usepackage{float}

\usepackage{times}
\usepackage{latexsym}
\usepackage{amsmath}
\usepackage{array}

\usepackage[T1]{fontenc}

\usepackage[utf8]{inputenc}

\usepackage{microtype}

\usepackage{inconsolata}

\usepackage{graphicx}

\usepackage{listings}
\usepackage{xcolor}
\usepackage{hyperref}
\usepackage{caption}

\definecolor{codegreen}{rgb}{0,0.6,0}
\definecolor{codegray}{rgb}{0.5,0.5,0.5}
\definecolor{codepurple}{rgb}{0.58,0,0.82}
\definecolor{backcolour}{rgb}{0.95,0.95,0.92}

\lstdefinestyle{mystyle}{
    backgroundcolor=\color{backcolour},   
    commentstyle=\color{codegreen},
    keywordstyle=\color{magenta},
    numberstyle=\tiny\color{codegray},
    stringstyle=\color{codepurple},
    basicstyle=\ttfamily\footnotesize,
    breakatwhitespace=false,         
    breaklines=true,                 
    captionpos=b,                    
    keepspaces=true,                 
    numbers=left,                    
    numbersep=5pt,                  
    showspaces=false,                
    showstringspaces=false,
    showtabs=false,                  
    tabsize=2
}

\usepackage{cleveref}

\title{Unlock the Correlation between Supervised Fine-Tuning and Reinforcement Learning in Training Code Large Language Models}
\vspace{-4em}

\author{Jie Chen\\
  ByteDance Seed\\
  \texttt{chenjiexjtu@gmail.com} 
  \And
  Xintian Han \\ 
  ByteDance Seed \\
  \texttt{hanxintian@bytedance.com} 
  \And
  Yu Ma \\
  ByteDance Seed \\
  \texttt{mayu.1231@bytedance.com}\\
  \AND
  Xun Zhou \\
  ByteDance Seed \\
  \texttt{zhouxun@bytedance.com}\\
  \And
  Liang Xiang \\
  ByteDance Seed \\
  \texttt{xiangliang@bytedance.com}
  }

\begin{document}
\maketitle
\begin{abstract}
Automatic code generation has been a longstanding research topic. With the advancement of general-purpose large language models (LLMs), the ability to code stands out as one important measure to the model's reasoning performance. Usually, a two-stage training paradigm is implemented to obtain a Code LLM, namely the pretraining and the fine-tuning. Within the fine-tuning, supervised fine-tuning (SFT), and reinforcement learning (RL) are often used to improve the model's zero-shot ability. A large number of work has been conducted to improve the model's performance on code-related benchmarks with either modifications to the algorithm or refinement of the dataset. However, we still lack a deep insight into the correlation between SFT and RL. For instance, what kind of dataset should be used to ensure generalization, or what if we abandon the SFT phase in fine-tuning. In this work, we make an attempt to understand the correlation between SFT and RL. To facilitate our research, we manually craft 100 basis python functions, called atomic functions, and then a synthesizing pipeline is deployed to create a large number of synthetic functions on top of the atomic ones. In this manner, we ensure that the train and test sets remain distinct, preventing data contamination. Through comprehensive ablation study, we find: (1) Both atomic and synthetic functions are indispensable for SFT's generalization, and only a handful of synthetic functions are adequate; (2) Through RL, the SFT's generalization to target domain can be greatly enhanced, even with the same training prompts; (3) Training RL from scratch can alleviate the over-fitting issue introduced in the SFT phase.
\end{abstract}

\section{Introduction}

Code generation based on natural language descriptions is a longstanding challenge \cite {summers1977methodology, liang2010learning, yin2017syntactic}, which is greatly powered by the recent development of Large Language Models (LLMs). Following rather early works such as Codex \cite {codex}, which is fine-tuned from the GPT-3 \cite{gpt3} model family  with publicly available code data from GitHub, a series of powerful Code LLMs have emerged in both open-source and proprietary domains~\cite{achiam2023gpt, team2023gemini, lozhkov2024starcoder, guo2024deepseek, codellama, pinnaparaju2024stable}. By deliberately designing data collection pipelines and training LLMs with vast code-related corpus, Code LLMs have shown superior performance in code generation tasks~\cite{guo2024deepseek, lozhkov2024starcoder, codellama}. However, as pointed in \citet {instruct-gpt}, making language models bigger does not inherently make them better at following a user's intent, which raises the importance of instruction-tuning or alignment \cite{askell2021generalalignment,bai2022alignment,zhou2024lima}. This has been repeatedly indicated by the performance gap between the base version and instruction-following version of various model series \cite{guo2024deepseek, codellama, li2023starcoder}.

Mostly, we can consider instruction-tuning through two approaches, supervised fine-tuning (SFT) and reinforcement learning (RL). For instance, both DeepSeekCoder and Code Llama utilized SFT to enhance their models' instruction-following and safety properties following their corresponding pretrained version. SFT relies on appropriate prompt-response pairs to make models follow user's intent and generate helpful, trustworthy and harmless responses, it is direct to implement but with easy tendencies to overfit.  Therefore, RL is usually introduced after SFT and benefits more from exploration and being able to interacting with the environment dynamically with diverse preferential reward signals. In the context of code generation, one of the most significant reward signals that should be taken into account is the feedback from unit tests of the generated program. For example, \citet{le2022coderl} used an offline actor-critic RL approach with unit test signals to further optimize a fine-tuned CodeT5 model \cite{wang2021codet5}. \citet {liu2023rltf} proposes an online RL framework to exploit the real-time unit test feedback and further categories the feedback signals into more fine-grained levels and provides varying rewards based on the ratio of passed test cases. Although both SFT and RL approaches have shown their effectiveness in alignment, there is still lack of thorough discussions on the correlation between SFT and RL in training Code LLMs, in either data or model aspect.

In this paper, by constructing a rather simple but effectively designed dataset for both SFT and RL, we provide new insights into how SFT and RL are correlated. More specifically, we manually designed a set of atomic functions, which is further used to generate composite datasets for SFT and RL training. Experimenting with these synthesized data, we can avoid issues like data contamination and focus on the ablation study. More importantly, composing the synthetic data from a controlled set of atomic functions in an organized order gives us a clear sight of analyzing how SFT and RL trained with different datasets affect the model's final generalization capability. In this paper, we would not only show the effects of different combined and augmented datasets on each stage separately, but would also show how SFT trained on different datasets would affect the following RL performance. In addition to testing on our synthesized datasets, we also evaluate our fine-tuned models on benchmarks like HumanEval \cite{codex} and MBPP \cite{mbpp}.

To summarize, the main contributions of this paper can be described as follows: (1) We propose an efficient method to synthesize python-function-like dataset, which is able to generate a large number of complex while legal python functions at a very low cost and has the potential to be extended to production environment; (2) With the manually crafted atomic functions and the synthetic ones, we discover that both datasets are indispensable for improving SFT's generalization ability to the target domain; (3) Specifically, we find just a handful of synthetic functions are adequate in guaranteeing the SFT's generalization performance. This indicates that a remarkable reduction to the cost of curating SFT dataset might be achieved, as in real-world scenarios, collecting or annotating synthetic functions often requires much more effort compared to their atomic counterparts; (4) Using the same training prompts for both SFT and RL, we show that the RL phase can still improve the model's generalization ability to the target domain by a large margin when initialized from the SFT's checkpoint. (5) We find the SFT model can easily overfit to the target domain; (6) When we run the RL process from scratch, a significant performance drop in the target domain can be observed, but the overfitting issue is well alleviated.

\section{Related work}

Following the rapid development of general large language models \cite {achiam2023gpt, team2023gemini, llama2, jiang2023mistral}, efforts are made for the release of Code LLMs as well \cite{lozhkov2024starcoder, guo2024deepseek, codellama}. Many works focus on different aspects of training Code LLMs such as curation of pretrain dataset \cite{li2023starcoder, lozhkov2024starcoder}, exploration of different fine-tuning methods \cite{luo2023wizardcoder, ppocoder}, evaluation benchmarks \cite{lai2023ds, mbpp}, etc. Due to the limitations of space, more detailed discussion on related works is shown in \cref{sec:related_work}.

\begin{figure*}[t]
\centering
\includegraphics[width=\textwidth]{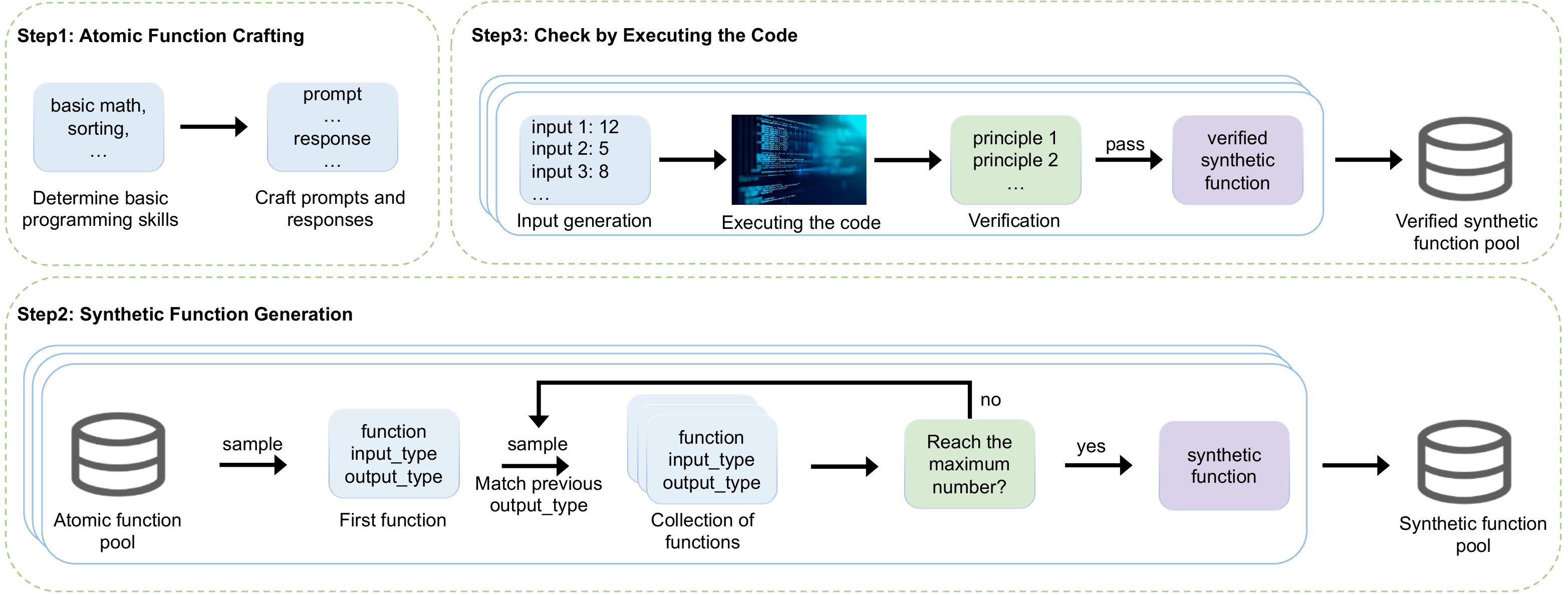}
\caption{Workflow of the data synthesizing process. In step 1, we manually craft 100 atomic python functions covering a few programming topics, such as math and sorting algorithms. In the second step, we run the data synthesizing pipeline and obtain a large number of synthetic functions. Finally, all the synthetic functions will be checked via the code sandbox for validity.}
\label{fig:generation}
\end{figure*}

\section{Methods}

Supervised Fine-Tuning (SFT) and Reinforcement Learning (RL) are usually performed after the pretraining stage. The motivation of SFT is considered to improve the model's zero-shot generalization capability \cite{wei2021finetuned},  while RL is often implemented to optimize the model's performance over certain objectives, e.g., the preference reward model. Usually, the RL process initializes from an SFT checkpoint and continues training on its own dataset. We can often observe that different datasets and SFT checkpoints pose a significant change to the performance of the RL process. However, we still lack a solid understanding about the correlation between SFT and RL. For instance, can we use the same prompts for both SFT and RL? What will happen if we train the RL without initializing from an SFT checkpoint?

In order to obtain a deep insight into the correlation between SFT and RL, we manually design a set of python atomic functions, then we propose an effective and efficient data synthesizing method which randomly draws a few functions from the atomic set and organizes the functions in reasonable orders to formulate a large amount of much more complicated but legal python functions. Experimenting with the synthesized data, we can avoid issues like data contamination and focus on the ablation study. We also implement the SFT and RL algorithms, where in the RL phase we use Proximal Policy Optimization (PPO) \cite{schulman2017proximal} to optimize the model. A code sandbox environment is devised to run the generated code snippet. The sandbox returns 1 if the code passes all the unit tests otherwise 0. Finally, we carefully design a number of ablation experiments to investigate the correlation between SFT and RL.

\subsection{Atomic Python Functions}

Ten programming topics have been designed, as displayed in~\cref{fig:atomic_functions}. Based on these topics, we hand-write 100 coding questions and invite a few experts to create responses for these questions using Python. We name these functions, the atomic python functions, since they form the basis of our data synthesizing process. Each atomic function has a prompt, namely the handwritten question, and a standard answer to the prompt and can be attributed to one of the ten programming topics,  e.g., string operation or math. 

One example of the atomic functions is shown in~\cref{fig:atom_exp}. We include the types of both input and output in the definition of the functions. In this work, to simplify the problem, we limit the input to four different types only: Int, String, List[Int], and List[String].

\subsection{Data Synthesizing}
We create the synthetic functions by concatenating a few atomic functions sequentially. One atomic function's output becomes another one's input. Therefore, the input type of the proceeding function has to be the same as the output type of the current one. The generation process is as follows: (1) Sampling the first one atomic function randomly; (2) Randomly selecting one from those atomic functions whose input type matches the previous function's output type; (3) Repeatedly running step 2 until the total number of functions reaches the desired count (in this work, the desired count is set to be 2, 3, or 4).

With the sampled atomic functions, we need to reconstruct their prompts and responses to build the synthetic function. For the prompts, we simply change some wording according to Chinese grammar. As for responses, we make sure the proceeding function's input variable's name matches the current function's output. One example of the synthetic functions can be found in \cref{fig:sync_exp}.

\textbf{Check Validity of synthetic functions.} Now we already have a pool of synthetic functions. However, not all the synthetic functions can be executed successfully, besides some of them may be too easy or even meaningless. Therefore, we further filter the synthetic functions by mocking the inputs and executing the code. And then we create some principles to filter out easy or meaningless functions that may be harmful for training. The mocked inputs are sampled following~\cref{tab:sample_criteria}.


For each synthetic function, we mock ten different sets of inputs according to its pre-defined input type, and then we execute the atomic functions contained in it in order. According to the execution feedback, we filter out functions which: (1) has runtime error; (2) has any atomic function outputting \textbf{None}. A None output usually means that the execution of the function is meaningless; (3) has an atomic function whose input and output are the same, indicating that the function does not do anything useful; (4) has same final outputs for the ten different sets of inputs, which means the function may be too easy. We depict the data synthesizing process in \cref{fig:generation}. For those valid synthetic functions, we keep the mocked inputs and their corresponding outputs, and use the \textbf{assert} formula to construct unit test cases.

\subsection{Supervised Fine-Tuning (SFT)}
Given the manually crafted atomic python functions and the data synthesizing method, we create four different datasets, namely Atom\_base, Composite\_A, Composite\_B, and Composite\_C. Atom\_base includes all the atomic python functions. In order to generate the other three composite datasets, we firstly divide all the atomic python functions into three distinct sets, namely set\_A, set\_B, set\_C, then we run the data synthesizing method on each set seperately to obtain the corresponding composite dataset. In this manner, we can avoid issues like data contamination. We also perform deduplication to the composite datasets, such as exact string matching and MinHash \cite{shrivastava2014defense}. Finally, we get 5000 <prompt, response> pairs for both Composite\_A and Composite\_B, and 1000 for Composite\_C, where Atom\_base and Composite\_A are used by our SFT process and Composite\_C serves as the test dataset for both SFT and RL.

\subsection{Code Sandbox Environment}
For safety and efficiency reason, we design a code sandbox environment to run the code snippets generated by the model \cite{codex}. As described in the data synthesizing method, given a prompt, apart from the code we also generate the unit test cases. The code sandbox is used in both training and evaluation stages, and only if one generated code passes all the unit tests will the sandbox determine it as correct. As only code can be executed, we extract the code block contained in the model's response and send it to the sandbox.

\subsection{Reinforcement Learning (RL)}
  Reinforcement learning (RL) has been proven to be effective in further improving the coding capability of LLMs after SFT. In this work, we implement the PPO algorithm which is widely used in the alignment stage. The learning objective of PPO is:
  \begin{equation}
  \mathop{max}_{\theta} \mathrm{E} \left[ \mathrm{L}^{\mathrm{CLIP}}(\theta) \right]
  \end{equation}
  where \(\theta\) denotes parameters of the LLM, and \(\mathrm{L}^{\mathrm{CLIP}}(\theta)\) is the Clipped Surrogate Objective,
  \begin{equation}
  \begin{split}
  \mathrm{L}^{\mathrm{CLIP}}(\theta) = \mathrm{E}_t [& \min(\frac{\pi_{\theta}(a_t|s_t)}{\pi_{\theta_{old}}(a_t|s_t)} \cdot A_t, \\ 
  & clip(\frac{\pi_{\theta}(a_t|s_t)}{\pi_{\theta_{old}}(a_t|s_t)}, \\
  & 1-\epsilon, 1+\epsilon) \cdot A_t)] 
  \end{split}
  \end{equation}
  where \(\frac{\pi_{\theta}(a_t|s_t)}{\pi_{\theta_{old}}(a_t|s_t)}\) is the ratio between current and last models, \(A_t\) is the advantage, and \(\epsilon\) is the clip range. The reward of PPO includes the score returned by the code sandbox \(r\) and a KL penalty term to ensure the RL policy does not deviate too much from its initialization checkpoint,
  \begin{equation}
  \mathrm{R}=r-\beta \mathrm{D}_{\mathrm{KL}}[\pi_{\theta}(a_t|s_t) || \pi^{\mathrm{Init}}(a_t|s_t)]
  \end{equation}
  where \(\beta\) is a coefficient controlling the strength of the KL penalty, and \(\pi^{\mathrm{Init}}(a_t|s_t)\) is the initialization model. In order to stabilize training, we apply whitening to both the reward and the advantage estimation \cite{schulman2015high}.


\section{Experimental Results}
In this section, based on our designed atomic python functions, Atom\_base, and the composite functions, Composite\_A, Composite\_B, and Composite\_C, we make a comprehensive study on both SFT and RL. As Composite\_C is totally isolated from Composite\_A and Composite\_B, we use it as a fair out-of-distribution test dataset in the following experiments. We use a GPT2-like model \cite{radford2019language} with 1.5B parameters to perform the SFT and RL processes, where the hidden size is 2048, the number of attention heads is 16, and the number of hidden layers is 24.

\subsection{Supervised Fine-Tuning (SFT)}
\subsubsection{Composite\_A and Atom\_base}
As we have 5000 <prompt, response> pairs in Composite\_A, but only 100 pairs are available in Atom\_base, in order to fairly compare the impact of Composite\_A and Atom\_base in the SFT phase, we conduct a few experiments by down-sampling Composite\_A or up-sampling Atom\_base.

\begin{figure}[t]
    \includegraphics[width=\columnwidth]{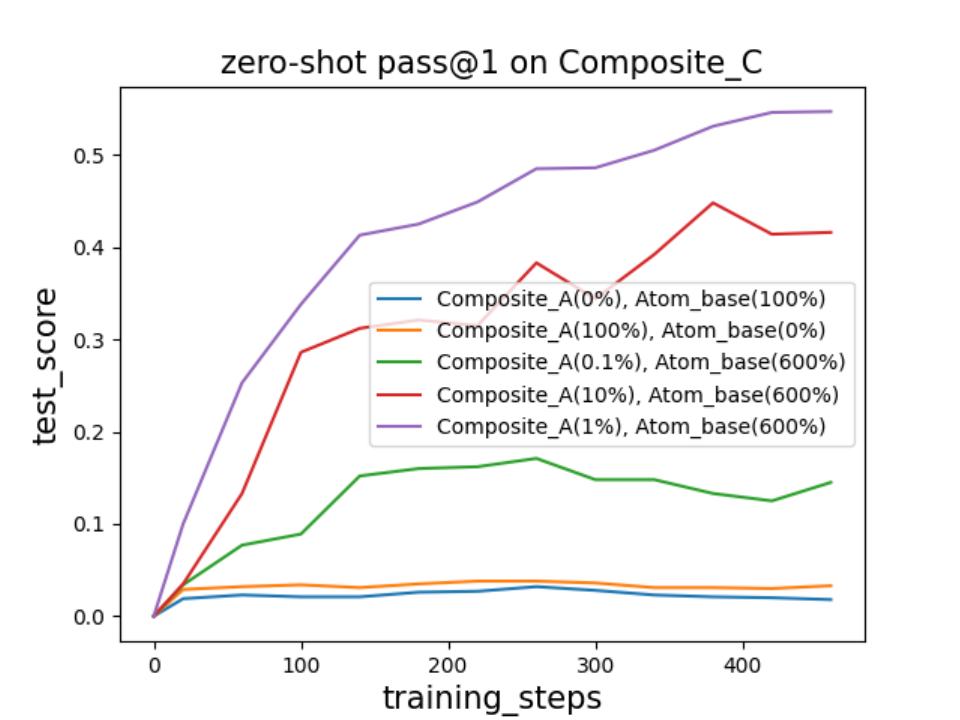}
    \caption{SFT with different mixtures of Composite\_A and Atom\_base, where we randomly down-sample Composite\_A while up-sample Atom\_base.}
    \label{fig:mixtures of Composite_A and Atom_base}
\end{figure}

\begin{figure}[t]
    \includegraphics[width=\columnwidth]{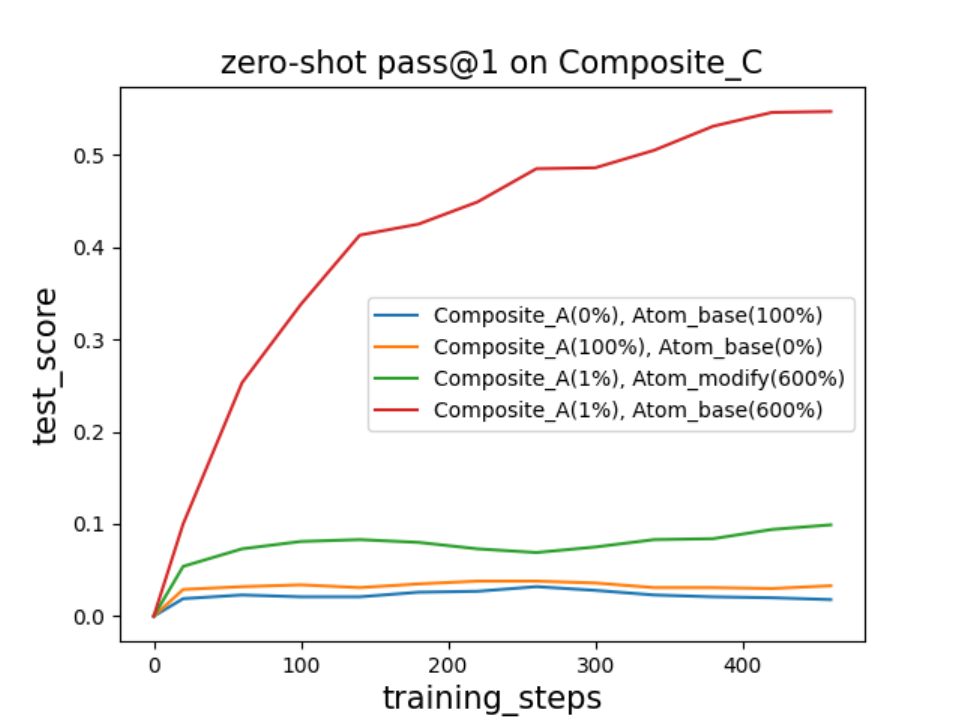}
    \caption{SFT with a mixture of Composite\_A and Atom\_modify. We randomly down-sample Composite\_A to 1\% while up-sample Atom\_modify to 600\%.}
    \label{fig:mixture of Composite_A and Atom_modify}
\end{figure}

In ~\cref{fig:mixtures of Composite_A and Atom_base}, we can see that the model fine-tuned on Atom\_base or Composite\_A only performs poorly on Composite\_C. By down-sampling Composite\_A and up-sampling Atom\_base, we find 'Composite\_A(1\%), Atom\_base(600\%)' performs the best, where only 50 <prompt, response> pairs are randomly drawn from Composite\_A, and achieves a 54.7\% zero-shot pass@1 on Composite\_C. As Composite\_A is constructed from a subset of Atom\_base with similar rules as Composite\_C, we attribute the contribution of Composite\_A to teaching the model how to use Atom\_base to solve a given synthetic problem, while Atom\_base endows the model with coding abilities at an atomic level. Given the experimental results, we can draw the conclusion: just a handful of composite data is good enough for the model to master the problem-solving skills in a specific domain, while enough atomic data is necessary.

\begin{figure}[t]
    \includegraphics[width=\columnwidth]{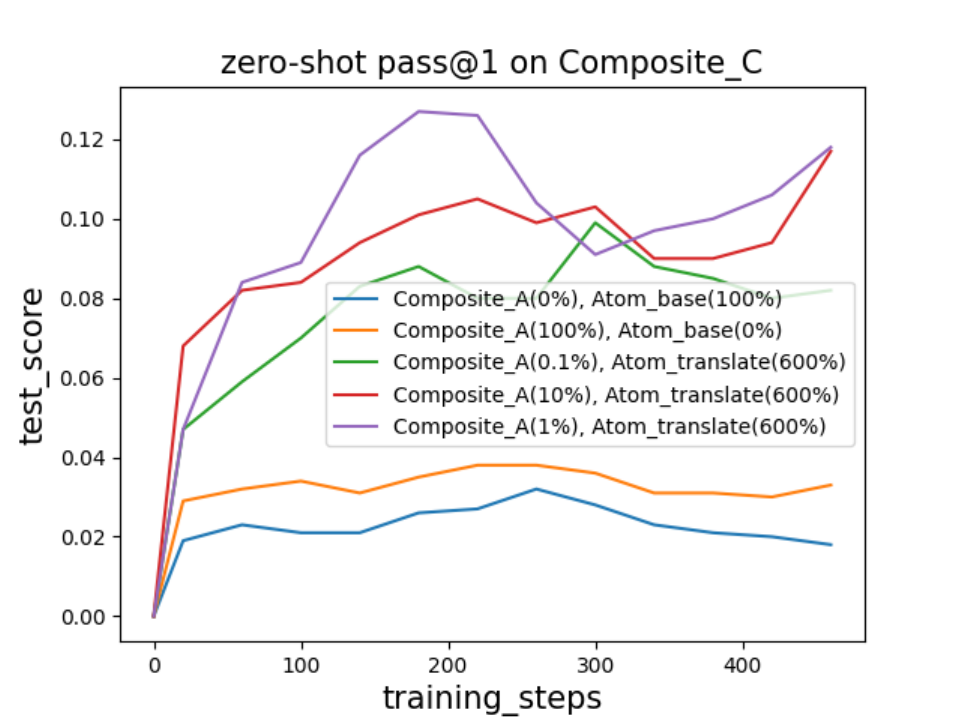}
    \caption{SFT with different mixtures of Composite\_A and Atom\_translate, where we randomly down-sample Composite\_A while up-sample Atom\_translate.}
    \label{fig:mixtures of Composite_A and Atom_translate}
\end{figure}

\begin{figure}[t]
    \includegraphics[width=\columnwidth]{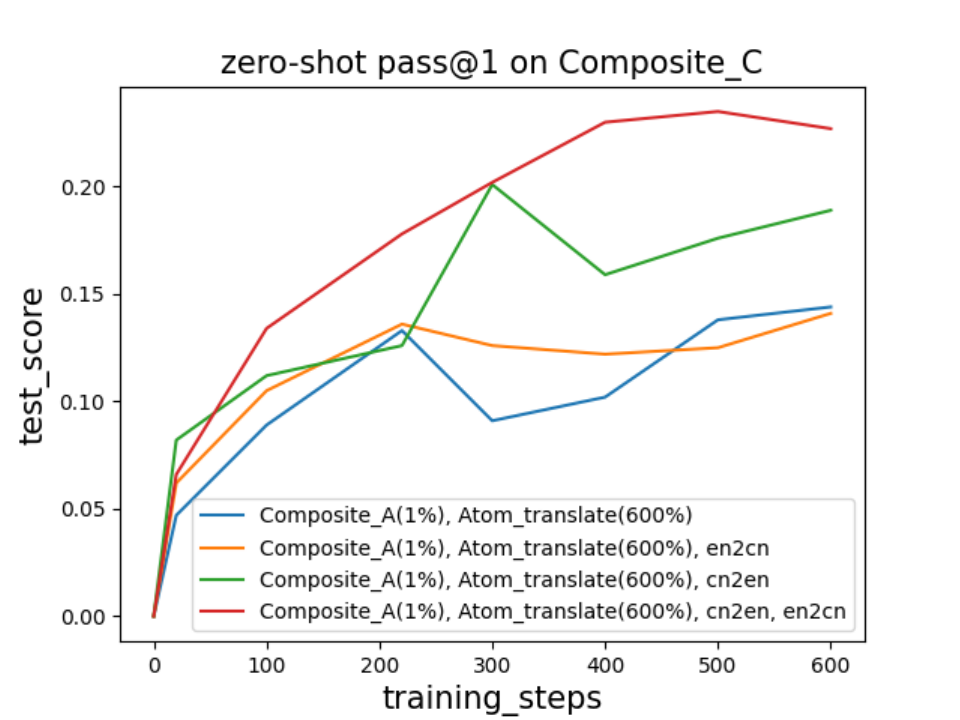}
    \caption{SFT with translated prompts, where en2cn denotes <prompt\_English, prompt\_Chinese> pairs, cn2en means <prompt\_Chinese, prompt\_English> pairs.}
    \label{fig:translated instructions}
\end{figure}

\subsubsection{Composite\_A and Atom\_modify}
As shown in~\cref{fig:mixtures of Composite_A and Atom_base}, enough atomic data is necessary for the model's generalization capability, we wonder what if we modify the code snippets and instructions a little bit in Atom\_base. We get a new atomic dataset, Atom\_modify, and a few examples are provided in ~\cref{fig:rewrite}.

In~\cref{fig:mixture of Composite_A and Atom_modify}, we can see that 'Composite\_A(1\%), Atom\_modify(600\%)' outperforms using Composite\_A or Atom\_base only, and achieves a 9.7\% zero-shot pass@1 on Composite\_C, but lags significantly behind 'Composite\_A(1\%), Atom\_base(600\%)'. It seems that even minor changes to the atomic data can lead to a large performance drop in generalization.

\begin{figure*}[t]
  \includegraphics[width=0.32\linewidth]{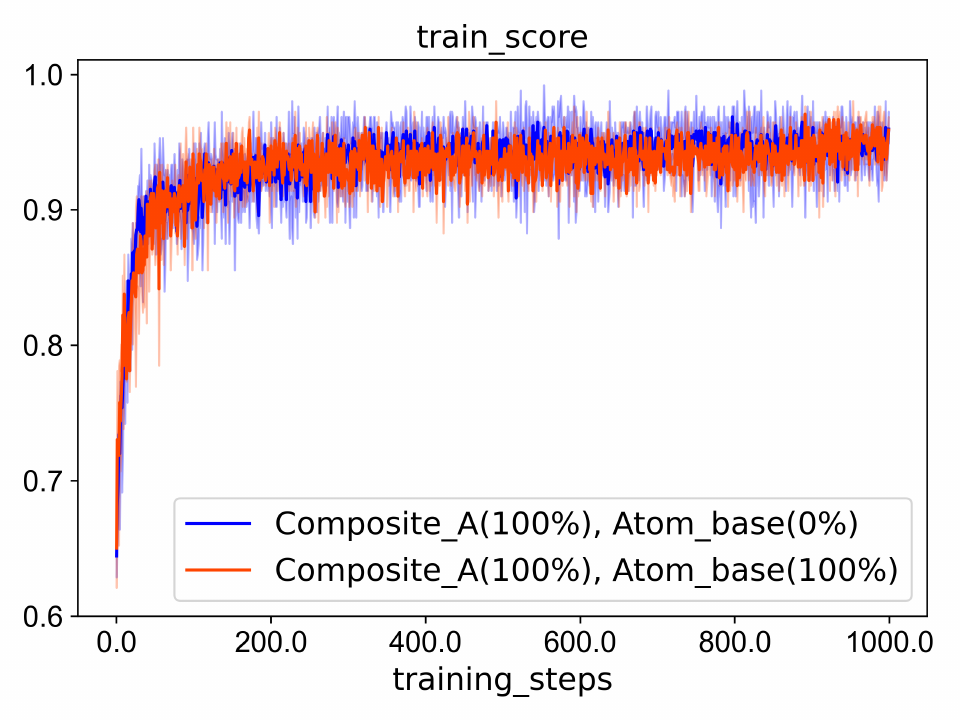} \hfill
  \includegraphics[width=0.32\linewidth]{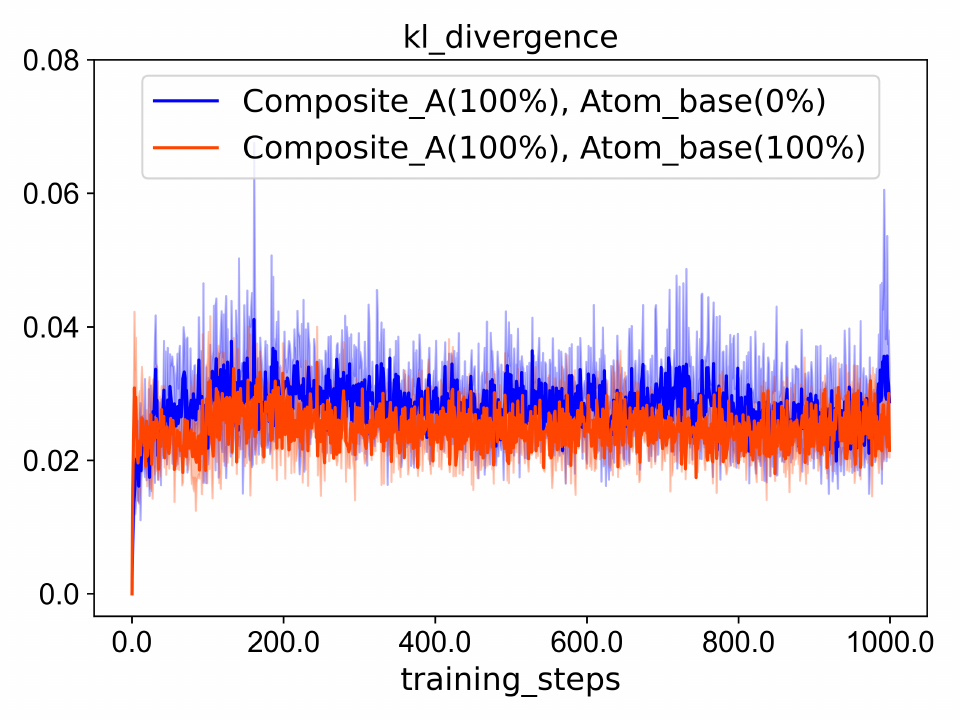} \hfill
  \includegraphics[width=0.32\linewidth]{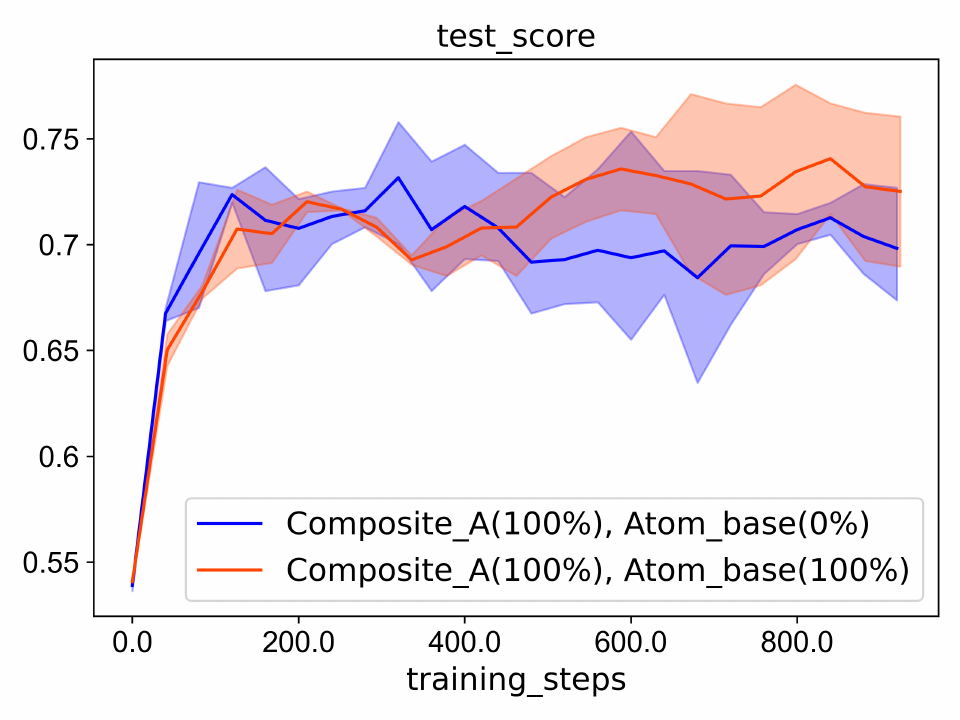}
  \caption {Continue to train the SFT model with RL on Composite\_A and Atom\_base. In training, we set the temperature to 0.2 and top-p to 0.7, while we use greedy decoding strategy for the testing.}
  \label{fig:RL1}
\end{figure*}

\begin{figure*}[t]
  \includegraphics[width=0.32\linewidth]{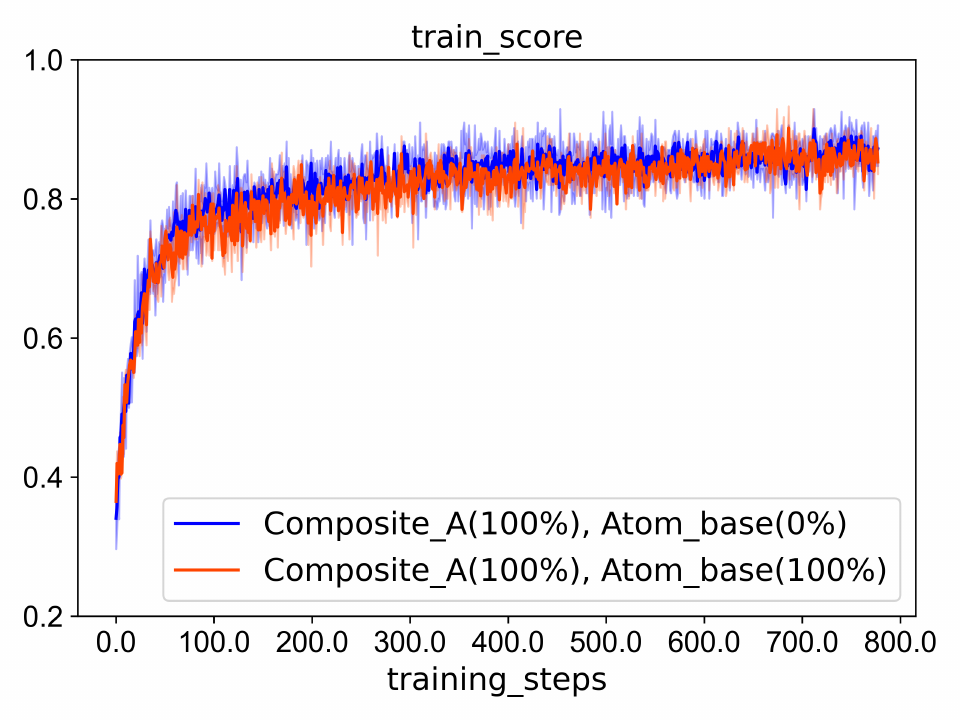} \hfill
  \includegraphics[width=0.32\linewidth]{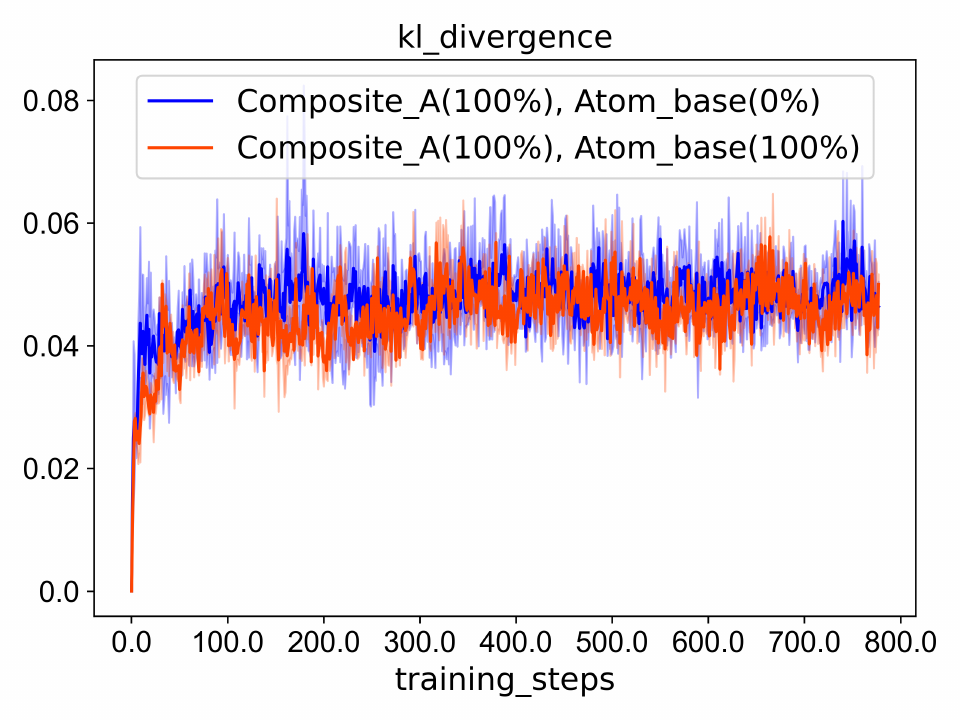} \hfill
  \includegraphics[width=0.32\linewidth]{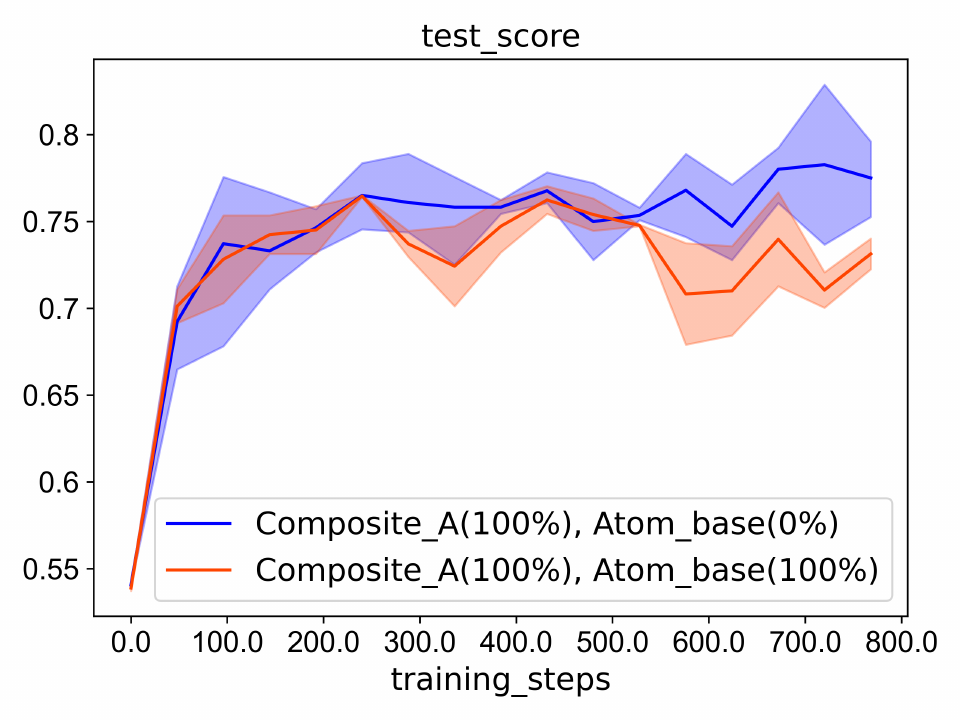}
  \caption {Continue to train the SFT model with RL on Composite\_B and Atom\_base. In training, we set the temperature to 0.2 and top-p to 0.7, while we use greedy decoding strategy for the testing.}
  \label{fig:RL2}
\end{figure*}

\subsubsection{Composite\_A and Atom\_translate}
In~\cref{fig:mixture of Composite_A and Atom_modify}, we find Atom\_modify under-performs Atom\_base significantly, we suspect it may be due to we rewrite both the prompts and responses in Atom\_base simultaneously. Therefore, in this section, we propose to modify only the prompts. In order to minimize the impact of our modification operation, we just translate the original Chinese prompts in Atom\_base into English, and get a new atomic dataset, Atom\_translate. Examples can be found in ~\cref{fig:translate}.

In~\cref{fig:mixtures of Composite_A and Atom_translate}, by mixing Composite\_A and Atom\_translate, the fine-tuned model outperforms using Composite\_A or Atom\_base only, and achieves a 12.6\% zero-shot pass@1 on Composite\_C. We note that the performance of 'Composite\_A(1\%), Atom\_translate(600\%)' is on par with that of 'Composite\_A(1\%), Atom\_modify(600\%)', which means even if we just translate the instructions, the model's generalization capability degrades significantly. We propose maybe the base model lacks the ability of English to Chinese understanding, therefore, we construct a new dataset containing <prompt\_Chinese, prompt\_English> and <prompt\_English, prompt\_Chinese> pairs, where prompt\_Chinese is the original instruction in Atom\_base, and prompt\_English denotes the translated version.

As shown in~\cref{fig:translated instructions}, the translated instructions do help the tuned model to generalize. However, when only the <prompt\_English, prompt\_Chinese> pairs are provided, the model failed to improve compared to 'Composite\_A(1\%), Atom\_translate(600\%)'. When both <prompt\_Chinese, prompt\_English> and <prompt\_English, prompt\_Chinese> pairs are available, the model benefits most from the instruction tuning phase by achieving a 23.5\% zero-shot pass@1 on Composite\_C, suggesting the importance of bi-directional language understanding.

\begin{figure*}[t]
  \includegraphics[width=0.32\linewidth]{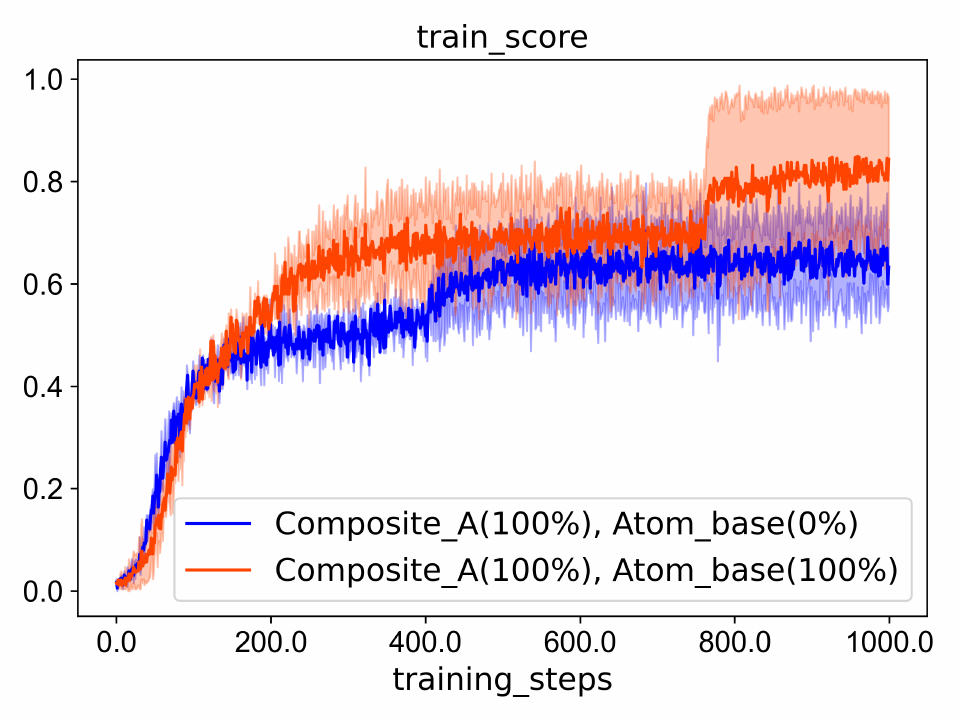} \hfill
  \includegraphics[width=0.32\linewidth]{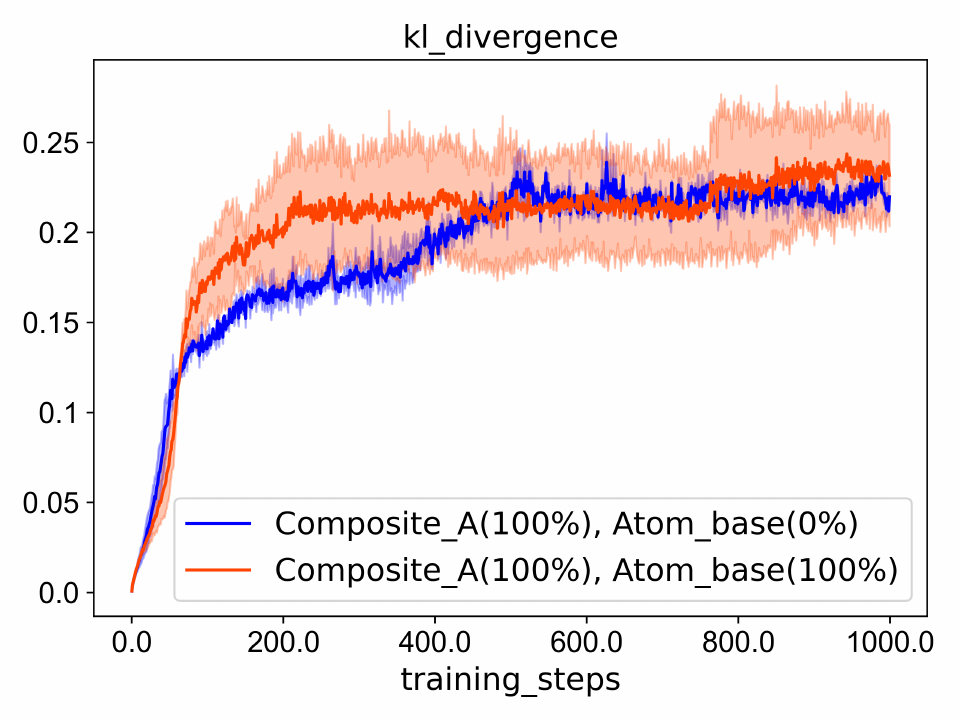} \hfill
  \includegraphics[width=0.32\linewidth]{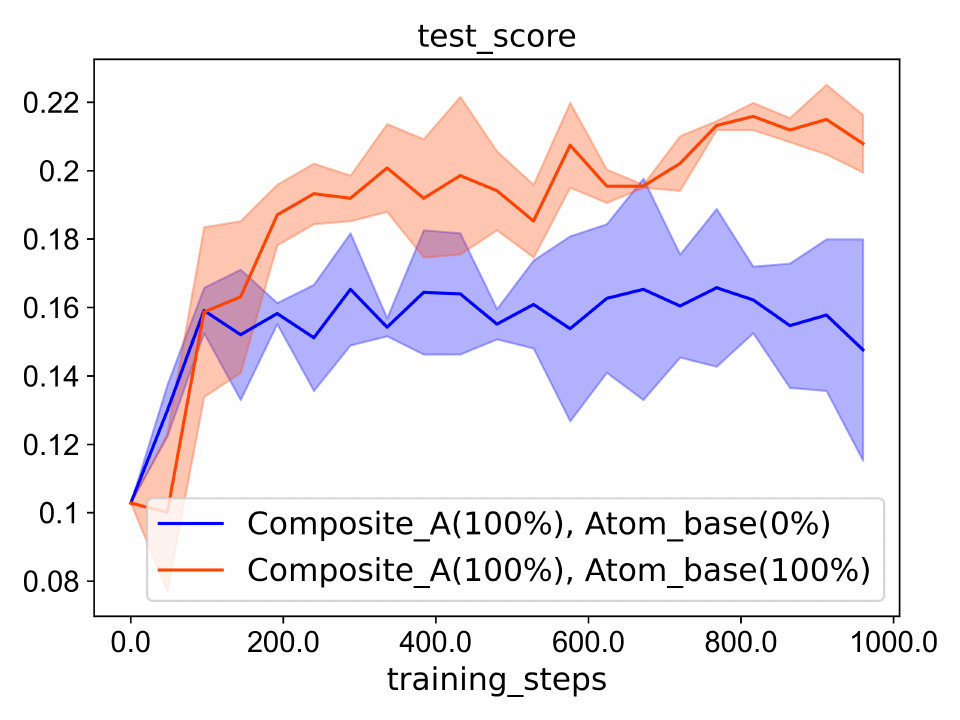}
  \caption {Train an untuned model with RL on Composite\_B and Atom\_base. In training, we set the temperature to 0.2 and top-p to 0.7, while we use greedy decoding strategy for the testing.}
  \label{fig:RL3}
\end{figure*}

\begin{figure*}[t]
  \includegraphics[width=0.32\linewidth]{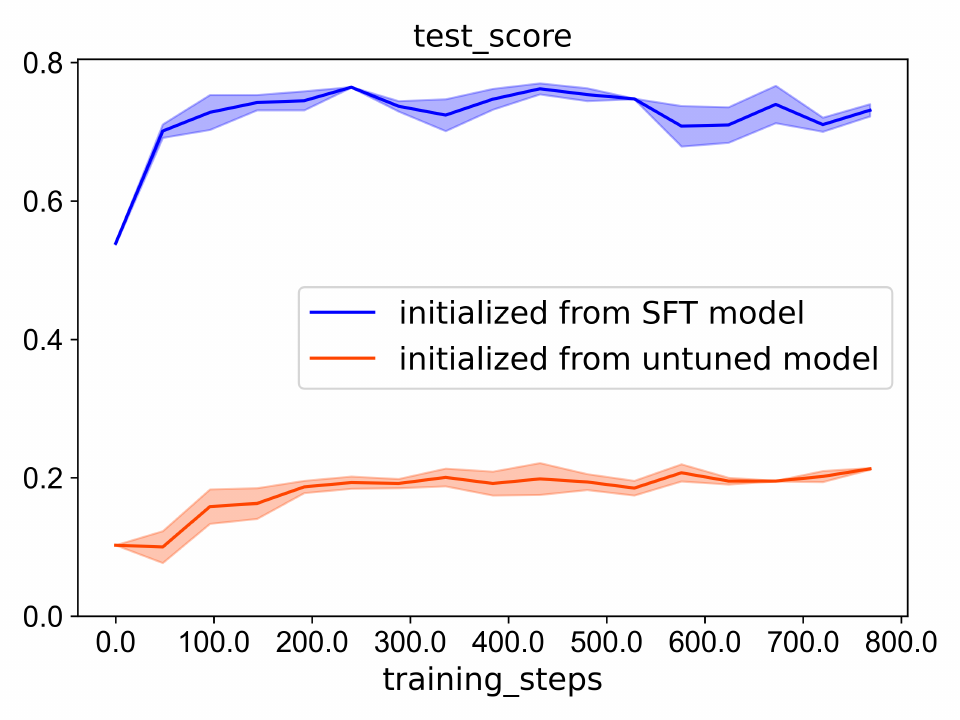} \hfill
  \includegraphics[width=0.32\linewidth]{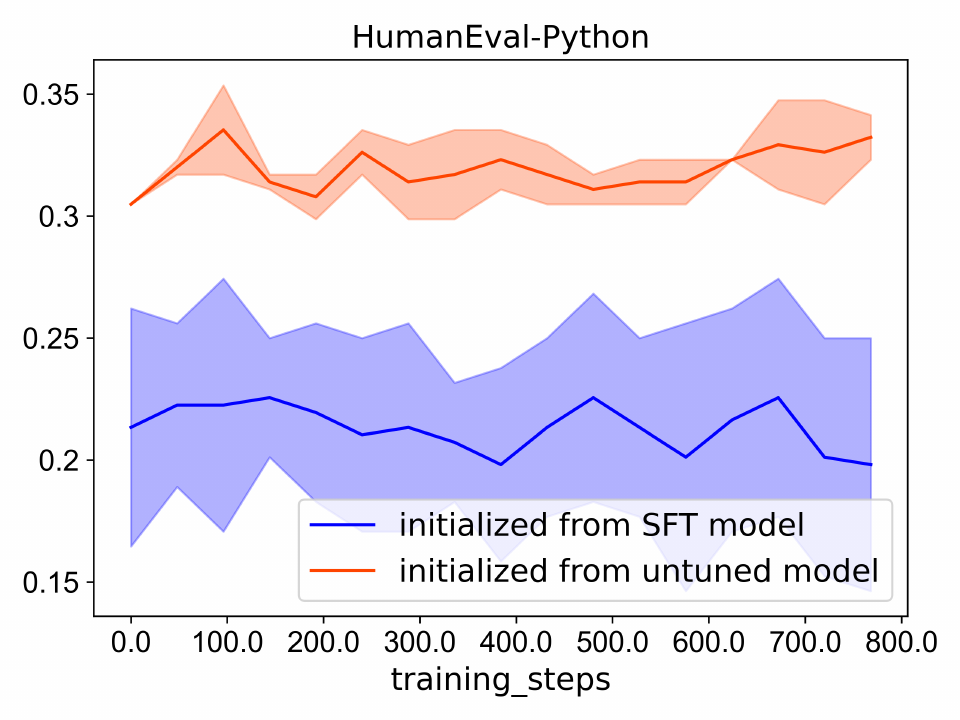} \hfill
  \includegraphics[width=0.32\linewidth]{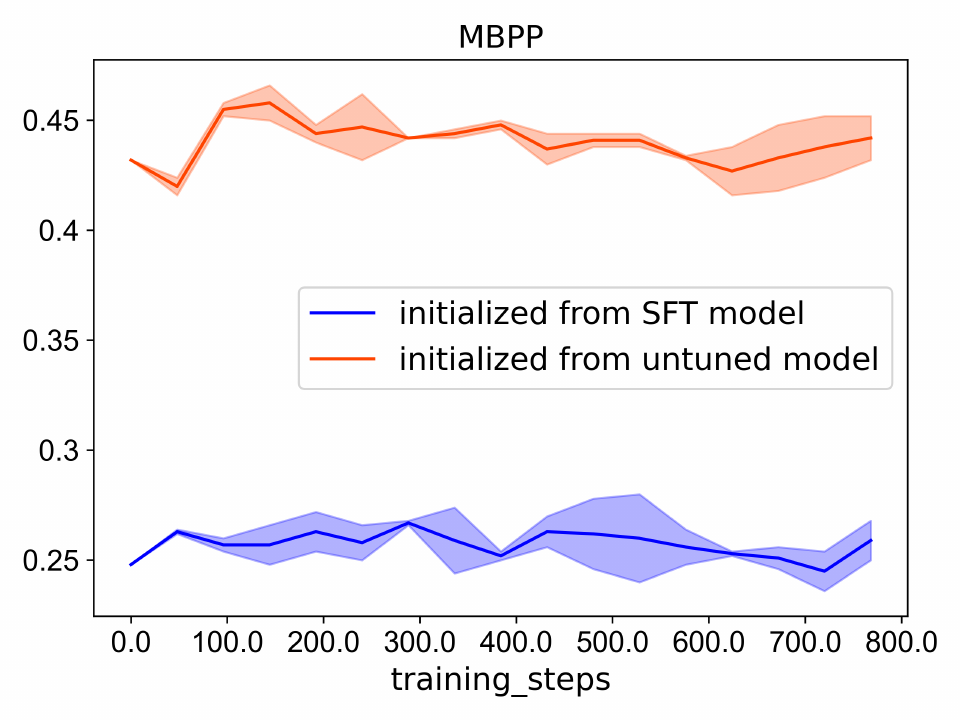}
  \caption {Comparisons between RL using different initialization models on Composite\_C (left), HumanEval-Python (middle), and MBPP (right). The training dataset is Composite\_B and Atom\_base, and we use greedy decoding strategy for the evaluation.}
  \label{fig:RL4}
\end{figure*}

\subsubsection{Summary}
Based on the experimental results of our SFT process, we can draw the following assumptions: (1) Both composite/synthetic and atomic data contribute to the generalization capability of tuned models; (2) A small number of composite data is adequate for the model to master problem solving skills; (3) Modification to the atomic data hurts the model's performance compared to using the original atomic data, while the model performs better with the modified atomic data than without it; (4) Bi-directional translation corpus is helpful in our setting.

\subsection{Reinforcement Learning}
Through SFT on Composite\_A and Atom\_base, the model do generalize well to Composite\_C and achieves a 54.7\% zero-shot pass@1. We wonder whether RL can further improve the model's generalization capability. Different from the data format used in the instruction tuning phase, namely <prompt, response>, we replace the response item with a set of unit tests, and the resulted data format is <prompt, unit\_tests>. A code sandbox environment is integrated into the training system to execute the model's generated code snippet, if the code passes all the unit test cases, the reward is 1.0 otherwise 0.0. 

As we use PPO to train our model, apart from the model itself, a reference policy model and a value model are added to the system. For the reference model, we initialize it from the instruction tuned checkpoint and keep its parameters frozen during training. For the value model, we choose to use a much smaller model than the policy model, where the hidden size is 1024, the number of attention heads is 16, and the number of hidden layers is 6. Please note that the value model is trained from scratch. Since we use a sandbox to calculate the reward, we avoid using a reward model and the so-called over-optimization issue.

\subsubsection{RL on Composite\_A and Atom\_base}
As shown in ~\cref{fig:mixtures of Composite_A and Atom_base}, after training on 'Composite\_A(1\%), Atom\_base(600\%)', the model achieves a 54.7\% zero-shot pass@1 on Composite\_C. We wonder what if we initialize from the tuned model, and continue to train the model on Composite\_A and Atom\_base with RL. Firstly, we continue to train the instruction tuned model on Composite\_A only, we can see that (~\cref{fig:RL1}) on the training dataset, namely Composite\_A, the model achieves a 95\% zero-shot pass@1, while on the test dataset, namely Composite\_C, the model achieves a 72.5\% zero-shot pass@1, meaning a 17.8\% increase on top of the SFT phase. Then, we perform RL with both Composite\_A and Atom\_base, as shown in~\cref{fig:RL1}, training on Composite\_A and Atom\_base achieves similar performance on the training dataset, however, it arrives at a higher pass rate on the test dataset, around 74\%.

\subsubsection{RL on Composite\_B and Atom\_base}
Composite\_A has been learned by the model in the SFT phase, so what if we train the reinforcement learning model with the unused dataset Composite\_B. Since Composite\_B has the same size as Composite\_A, we reproduce the experiments in last section to ablate the influence of different reinforcement learning datasets.

In~\cref{fig:RL2}, we use RL to optimize the SFT model on Composite\_B and Atom\_base. When trained on Composite\_B only, the model achieves a 88\% zero-shot pass@1 on the training dataset, which is a bit lower than Composite\_A's experiments. We think this is reasonable as Composite\_A's standard answers have been shown to the model during instruction tuning phase. However, it achieves a higher pass rate on the test dataset, around 77.5\%, which is a 2.5\% absolute increase compared with the best result in Composite\_A's experiments. When trained on both Composite\_B and Atom\_base, the model performs similarly on the training dataset, however, it performs a little bit worse on the test dataset.

\subsubsection{Initialized from an Untuned Model} \label{sec:rlfromuntuned}
In the above experiments, we arrive at the following insights: (1) Even with the same training prompts, RL can improve the model's generalization by a large margin over SFT; (2) Providing RL with unseen training prompts, the model performs better than using the same data as in SFT. In this section, we would like to investigate what if we train from an untuned model with RL, where "untuned" means we skip SFT on Composite\_A and Atom\_base. As depicted in~\cref{fig:RL3}, initialized from an untuned model, the RL performs better in both training and testing when using Composite\_B and Atom\_base than Composite\_B only. This is reasonable, as Composite\_B does not include the atomic functions used to construct Composite\_C. Compared to the results in ~\cref{fig:RL2}, we can conclude that, even if we use the same training dataset for reinforcement learning, performing SFT prior to RL does help the model to achieve a much higher generalization score in the target domain.

In addition to Composite\_C, we also include HumanEval-python and MBPP in the test dataset. As shown in~\cref{fig:RL4}, although the supervised fine-tuned model generalizes well to Composite\_C (~\cref{fig:RL3}), it performs poorly on HumanEval-python and MBPP, suggesting it already overfits to the composite dataset. However, when initializing reinforcement learning from the untuned model, it can sustain the model's performance on HumanEval-python and MBPP, and meanwhile improve its generalization score over Composite\_C. More training details can be found in ~\cref{fig:diff_init_HardB_full} and ~\cref{fig:diff_init_HardB_Atom_full}.

\subsubsection{Summary}
Based on the experimental results of our reinforcement learning process, we can draw the following assumptions: (1) Initialized from the SFT model, after optimization by RL with the same dataset, the model can still obtain a significant improvement on the test dataset; (2) Continuing training with RL of the SFT model on an unseen dataset, namely Composite\_B, performs better than with the same prompts used in instruction tuning; (3) Different initializations pose an significant impact to the reinforcement learning phase, and SFT prior to RL does help the model to achieve a much higher generalization score in the target domain; (4) Training RL from scratch can alleviate the over-fitting issue introduced in the SFT phase. 








\section{Conclusion}
We investigate the correlation between supervised fine-tuning (SFT) and reinforcement learning (RL) in training a code language model. To facilitate our research, an efficient data synthesizing pipeline is designed. The pipeline can generate a large number of complex while legal python functions at a very low cost, demonstrating the potential to be extended to production environment. Comprehensive ablation study has been performed for both SFT and RL. As for SFT, we discover that the atomic and the synthetic functions are all indispensable for improving generalization performance in the target domain. Specifically, we find just a handful of synthetic functions are adequate for guaranteeing the SFT's generalization ability. This indicates that a remarkable reduction to the cost of curating SFT dataset might be achieved, as in real-world scenarios, collecting or annotating synthetic functions often requires much more effort compared to their atomic counterparts. Through RL, the SFT's generalization to target domain can be greatly enhanced, even with the same training prompts. Moreover, we find training RL from scratch can alleviate the over-fitting issue introduced by the SFT phase.

In the near future, we are going to extend the insights obtained in this work by: (1) Instead of manually crafting the atomic functions, the large corpus of code pre-training data can be used to filter atomic functions; (2) Except for Python, more programming languages will be considered; (3) We are going to scale up both the model and the dataset and see whether the insights in this work still hold; (4) More code-related benchmarks will be included.



\section{Limitations}

The ablation study is performed on models with 1.5B parameters, we cannot assure that the obtained insights in this work still hold for larger models. And we only handcraft 100 atomic functions covering 10 programming topics to form the synthetic dataset, which apparently cannot represent the whole coding skill set. We limit the programming language to Python in this work, so the conclusions may not transfer to other languages such as C++ or Go.

We consider only the code generation and code completion tasks in this work, however, there exist many other code-related tasks. For instance, code understanding, code execution, code debugging, code translation, etc. The insights obtained may not naturally generalize to other tasks, and we need to include many other coding problems into our experiment. Moreover, we implement the PPO algorithm for the reinforcement learning stage. Given that a number of alignment techniques have been newly proposed, we may perform the designed experiments in this work with other algorithms. Besides the synthetic test dataset Composite\_C, we only include HumanEval-Python and MBPP as the additional test sets, which is far more than enough. We will evaluate the trained models on more code benchmarks in the future, such as DS-1000, LiveCodeBench, etc.


\appendix
\clearpage
\section{Related Work} \label{sec:related_work}

\subsection{Code LLMs}
Following the rapid development of general large language models \cite {achiam2023gpt, team2023gemini, llama2, jiang2023mistral}, Code LLMs \cite{lozhkov2024starcoder, guo2024deepseek, codellama} have emerged as well targeting specifically code-related tasks such as code generation, code understanding, code-related QA, etc. While code generation following natural language instructions is a primary focus up to this stage, Code LLMs have exhibited the potential to enhance all phases of the software development cycle \cite {lozhkov2024starcoder, codellm-survey}. Most of the Code LLMs such as AlphaCoder \cite{alphacoder}, StarCoder \cite{lozhkov2024starcoder}, DeepSeekCoder \cite{guo2024deepseek} are trained on code-corpus, while CodeLlama \cite{codellama} finds that initializing the model from LLaMA2 \cite{llama2} which is a foundation model pretrained on general-purpose text and code data outperforms the architecture trained on code only for a given budget. Although the intended goal of Code LLMs are far more comprehensive \cite{fan2023large,wang2024software}, at this stage, the most common way to evaluate the performance of Code LLMs is through several standard code-generation benchmarks \cite{codex, mbpp, lai2023ds, jain2024livecodebench} and evaluating the passed rates via pre-defined unit tests.

\subsection{Supervised Fine-Tuning for Code LLMs}
Supervised fine-tuning in the early stages was primarily used to enhance the cross-task generalization capabilities of language models by incorporating diverse NLP task instructions \cite{xu2023wizardlm, nlp-ft3, nlp-ft2, nlp-sft1}. While fine-tuning LMs with these close-domain NLP tasks has shown consistent performance improvements, in the era of LLMs, it often falls short for real user's varying intents. Therefore, various efforts have been made lately to fine-tune both general-purpose LLMs and Code LLMs by searching for the best ways of constructing open-domain instruction data \cite{wei2023magicoder, liu2023makes}. Vicuna \cite{chiang2023vicuna} utilized directly gathered user-shared conversations to fine-tune Llama. Wizard \cite{xu2023wizardlm, luo2023wizardcoder} instead introduced the Evol-Instruct method, which evolved existing instruction data to generate more complex and diverse datasets. CodeLlama primarily generated and selected their instruction fine-tuning data in a self-instruct way via prompting LLaMA2 for interview-style programming questions and then generating Python solutions and unit tests by prompting CodelLama 7B base model \cite{codellama}.



\subsection{Reinforcement Learning for Code LLMs}
Reinforcement learning essentially learns the optimal policy by obtaining reward signals from the real environment \cite{sutton1999policy}. Its role in fine-tuning LLMs is becoming more significant since \citet{instruct-gpt} demonstrates its promising effects on aligning the model with human intent by exploiting human feedback. With regard to Code LLMs, there has also been emerging research on utilizing RL to further enhance model performance. As mentioned above, one of the common ways to construct reward signals is to utilize unit test signals \cite{ppocoder, liu2023rltf}. In PPOCoder \cite{ppocoder}, it obtains the reward signal at the end of each generation episode based on the generated code's syntactic and functional correctness as well as its structural alignment with executable codes. And RLTF \cite{liu2023rltf} takes a small step further by adding additional fine-grained feedback signals that points out specific reasons for errors. Meanwhile, it also takes into account the number of passed test examples by introducing an "adaptive feedback" score. To address the exploration challenge in RL, StepCoder \cite{dou2024stepcoder} proposed breaking the long sequences code generation task into a curriculum of code completion subtasks to alleviate the complexities associated with exploration in RL. In this work, we utilizes the reward signal from executing the generated response against unit tests as well.

\clearpage
\section{Handcrafted Atomic Functions}


\begin{minipage}{0.49\textwidth}
\centering
\includegraphics[width=0.96\linewidth]{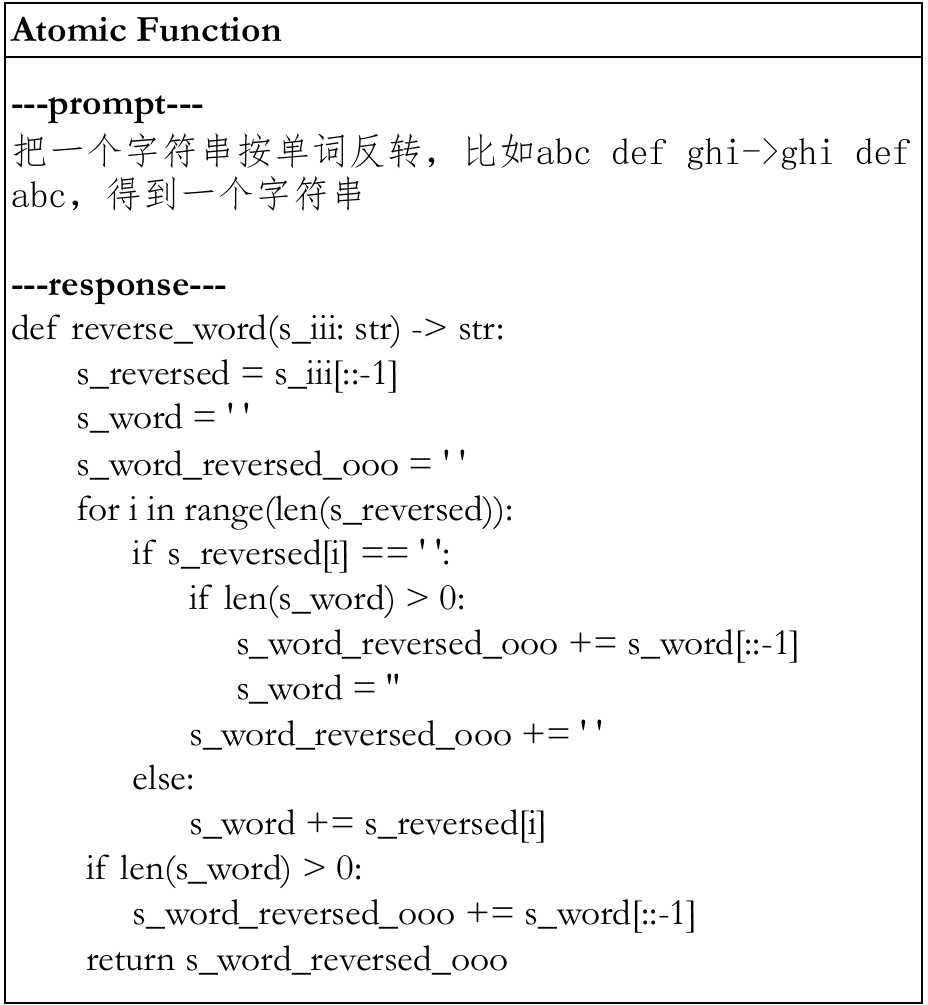}
\captionof{figure}{An example of the handcrafted atomic functions. We first design the prompt and then write the response. Please note that the prompts are written in Chinese, and the responses are programmed with Python.}
\label{fig:atom_exp}
\end{minipage}

\vfill
\vspace{20em}
\hfill
\hspace{20em}

\begin{minipage}{0.49\textwidth}
    \centering
    \vspace{2em}
    \begin{tabular}{|c|>{\centering\arraybackslash}p{5cm}|}
    \hline
         \textbf{Input Type} & \textbf{Sampling Criteria} \\
         \hline
         Int & Randomly sample an integer from 0 to 999. \\
         \hline
         String & Randomly sample a string from numbers, letters, spaces and a mixture of them. \\
         \hline
         List[Int] & Randomly sample the length of the list from 16 to 64,  then sample for each integer in the list from 0 to 999. \\
         \hline
         List[String] & Randomly sample the length of the list from 16 to 64, then sample for each item in the list from numbers, letters, spaces and a mixture of them. \\
         \hline
    \end{tabular}
    \captionof{table}{Sampling criteria in mocking inputs for the synthetic functions. In this work, we limit the input variable to four different types, namely Int, String, List[Int], List[String]. For Int, we set the minimum value to 0 and the maximum to 999. Moreover, the length of List is restricted between 16 and 64.}
    \label{tab:sample_criteria}
\end{minipage}

\vfill
\vspace{2em}
\hfill
\hspace{-70em}
\begin{minipage}{0.88\textwidth}
\includegraphics[width=1.0\linewidth]{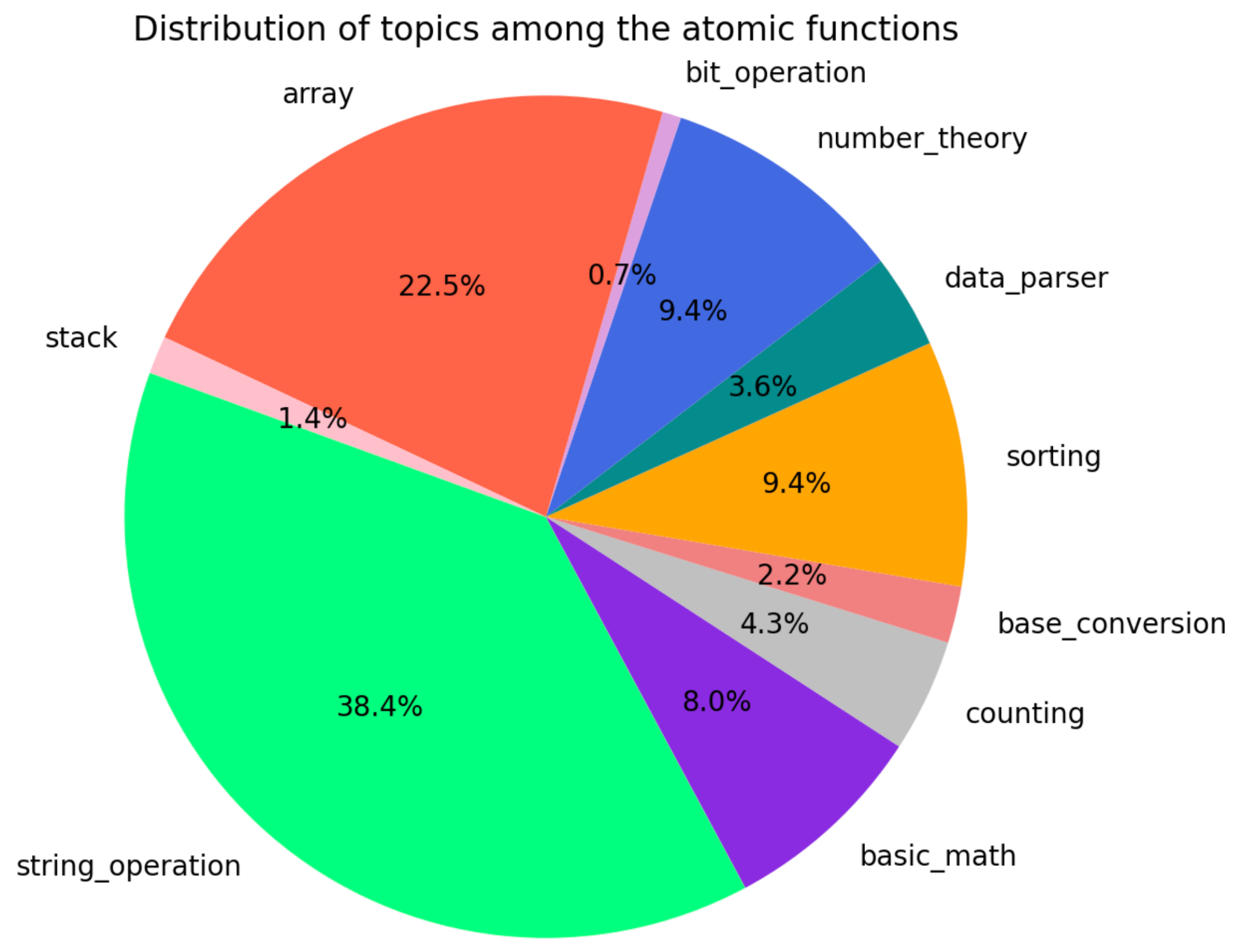}
\captionof{figure}{Distribution of programming topics among the handcrafted atomic python functions.}
\label{fig:atomic_functions}
\end{minipage}

\clearpage
\section{An example of synthetic functions}
\begin{minipage}{1.0\textwidth}
\strut\newline
\centering
\includegraphics[width=1.0\textwidth]{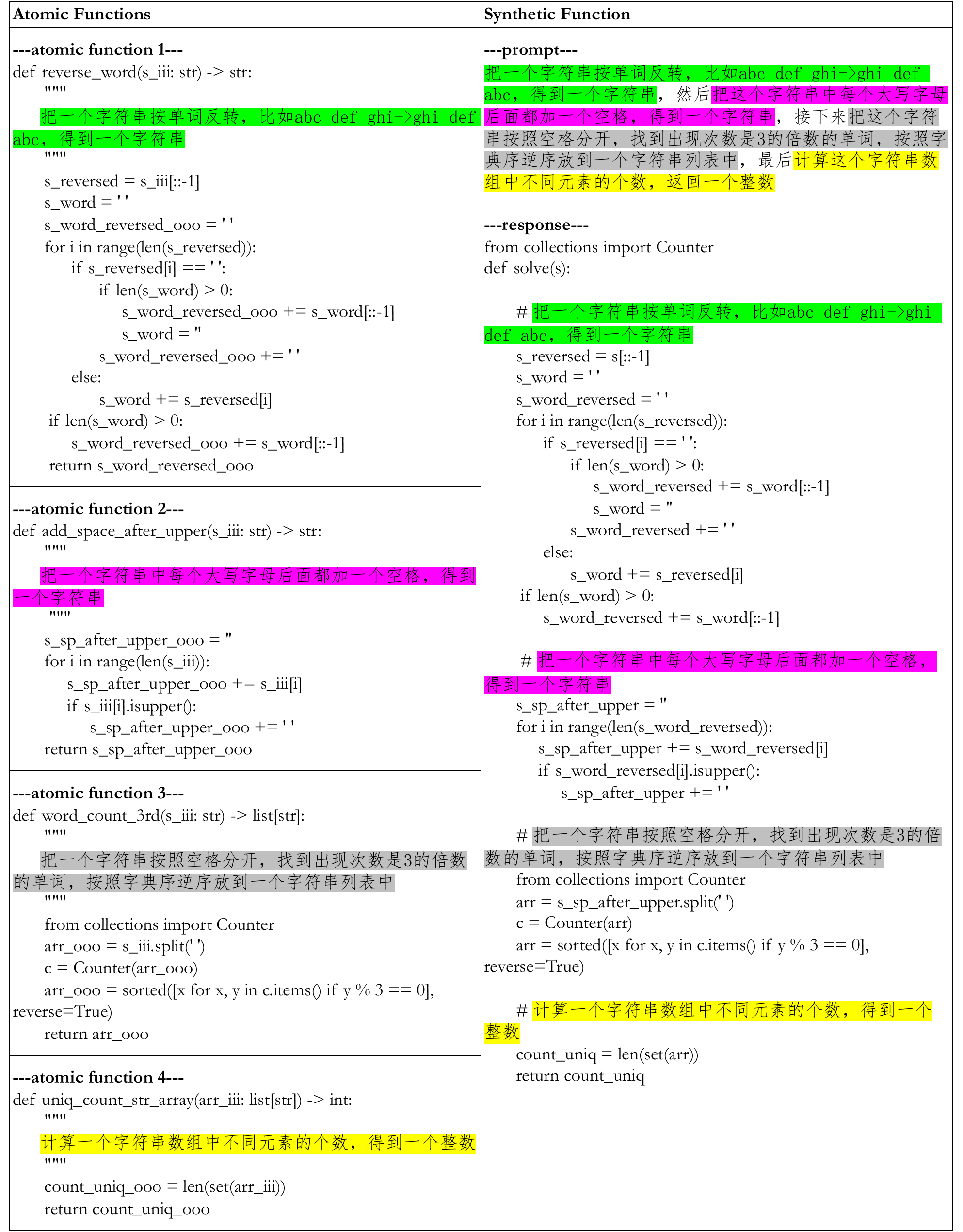}
\captionof{figure}{An example of synthetic functions, where on the left column are the sampled four atomic functions. We concatenate the atomic functions one by one to create the synthetic function. For sake of use, we name all the synthetic functions "solve". The validity of the synthetic functions is checked by the code sandbox.}\label{fig:sync_exp}
\end{minipage}

\clearpage
\section{Examples of modified and translated atomic functions}
\begin{minipage}{0.96\textwidth}
\strut\newline
\centering
\includegraphics[width=0.96\textwidth]{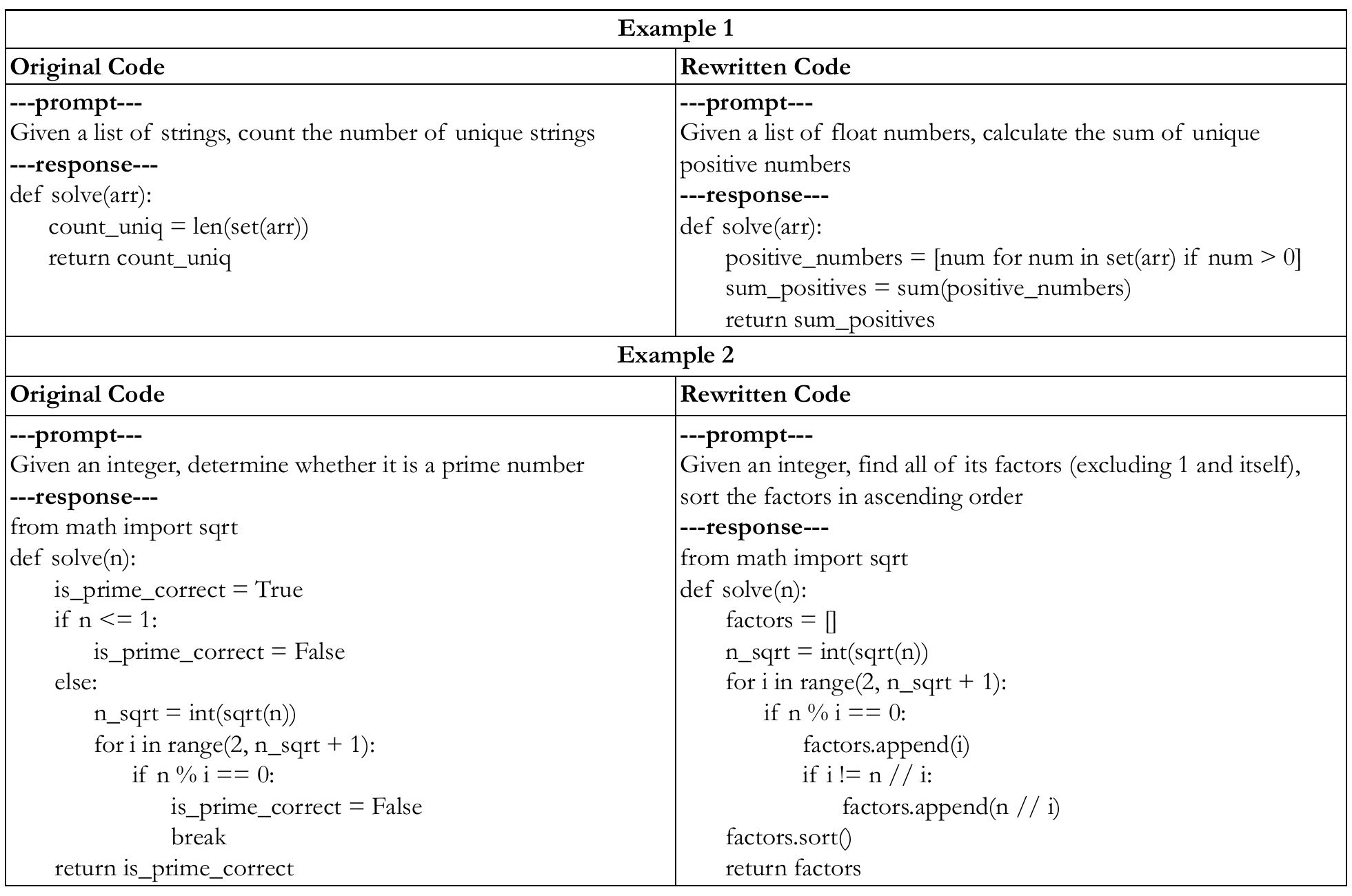}
\captionof{figure}{Two examples of the modified atomic functions, where on the left column are the original functions and the rewritten ones are placed on the right column. Please note that we translate the prompts into English here to ease understanding, while in the ablation experiment we keep using the Chinese version.}\label{fig:rewrite}
\strut\newline
\includegraphics[width=0.96\textwidth]{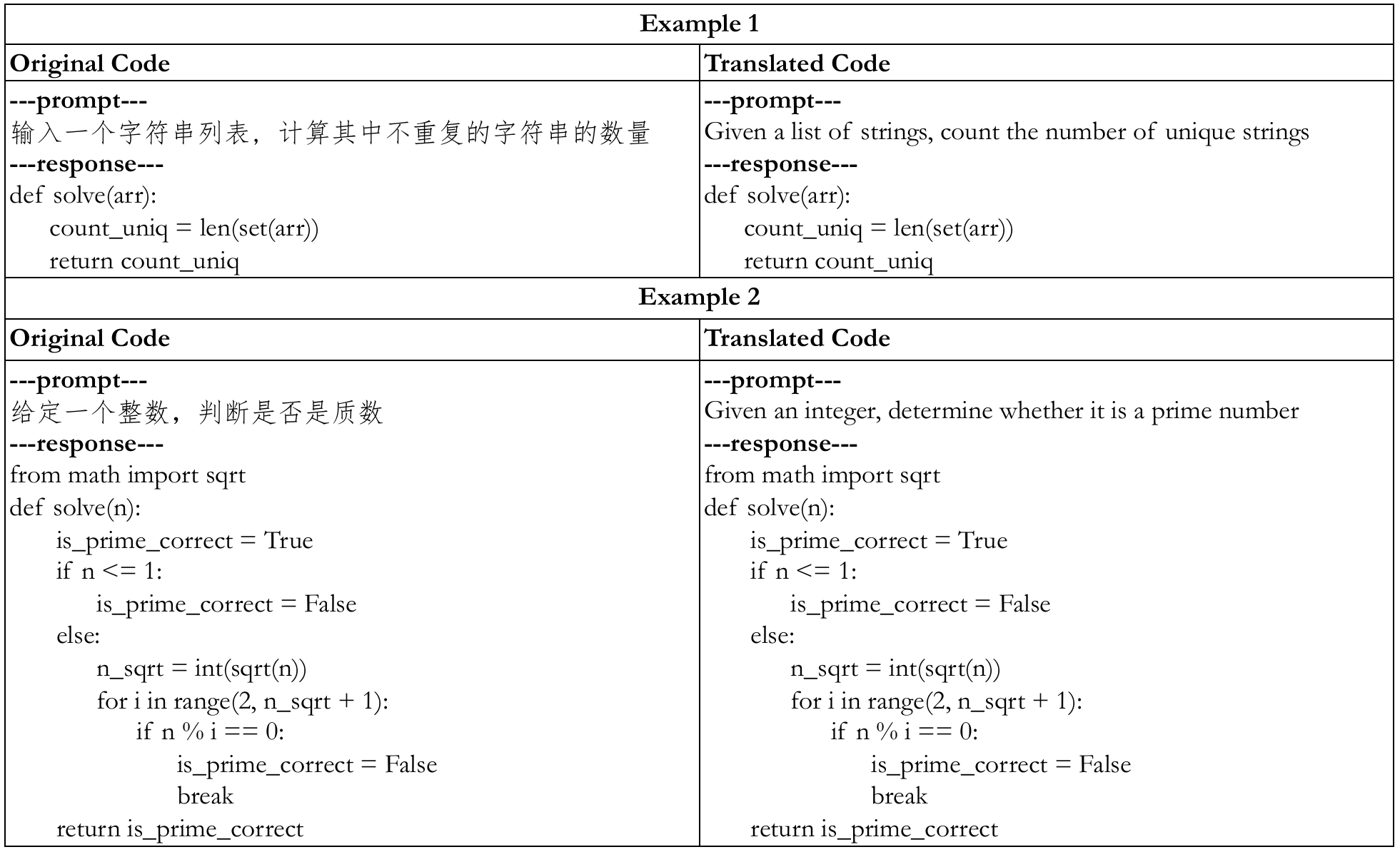}
\captionof{figure}{Two examples of the translated atomic functions, where on the left column are the original functions prompted in Chinese. We manually translate the Chinese prompts into English.}\label{fig:translate}
\end{minipage}
 
\clearpage
\section{RL from Different Initializations}
\begin{minipage}{0.96\textwidth}
  \includegraphics[width=0.48\linewidth]{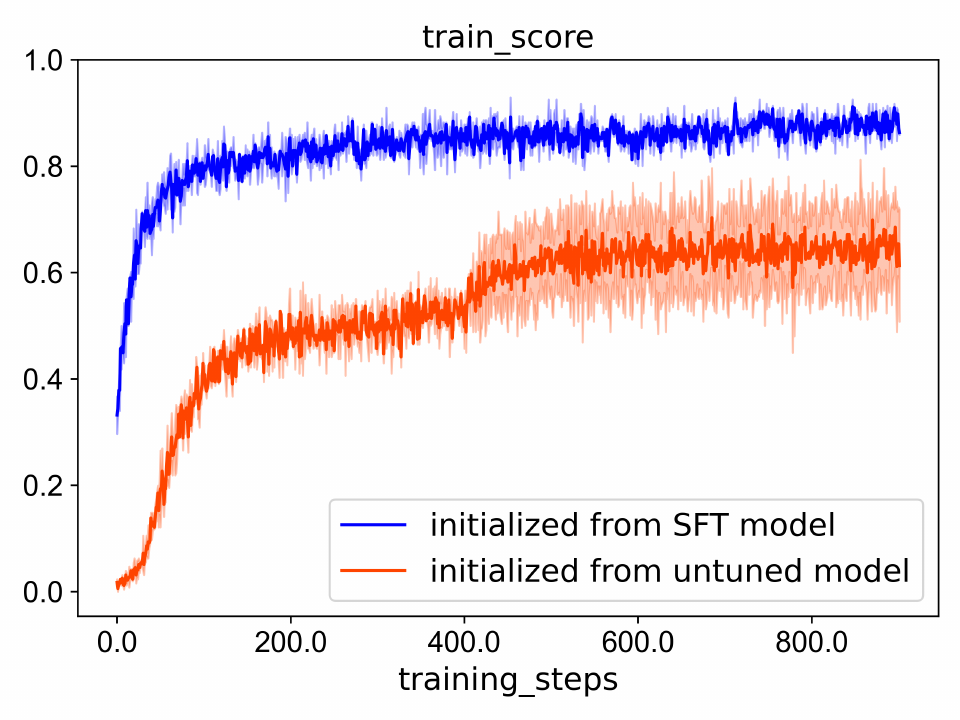} \hfill
  \includegraphics[width=0.48\linewidth]{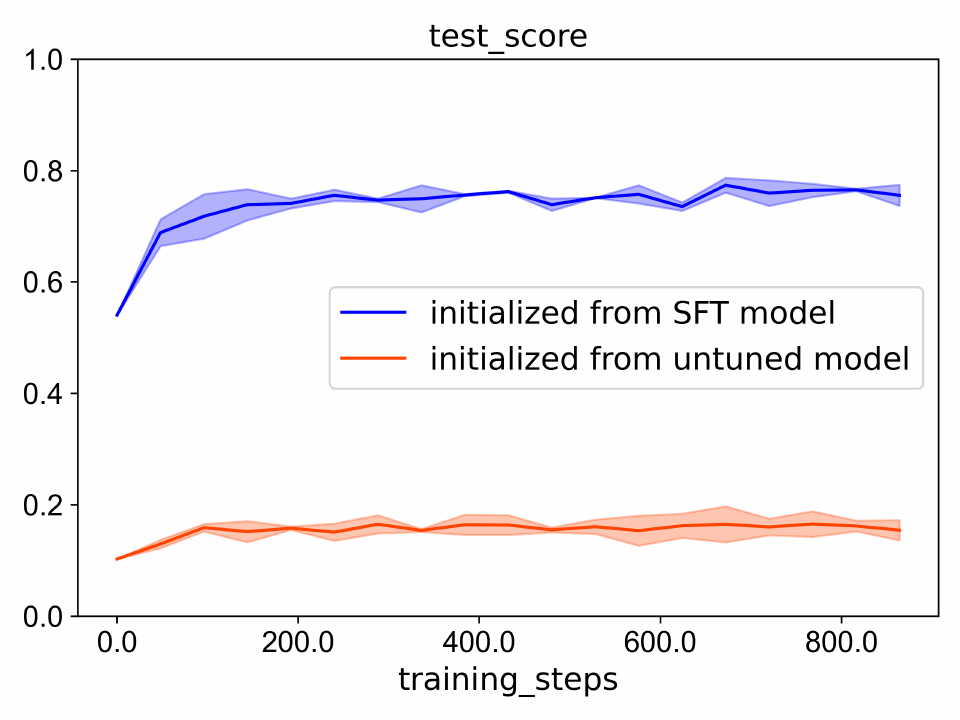}
  \quad
  \includegraphics[width=0.48\linewidth]{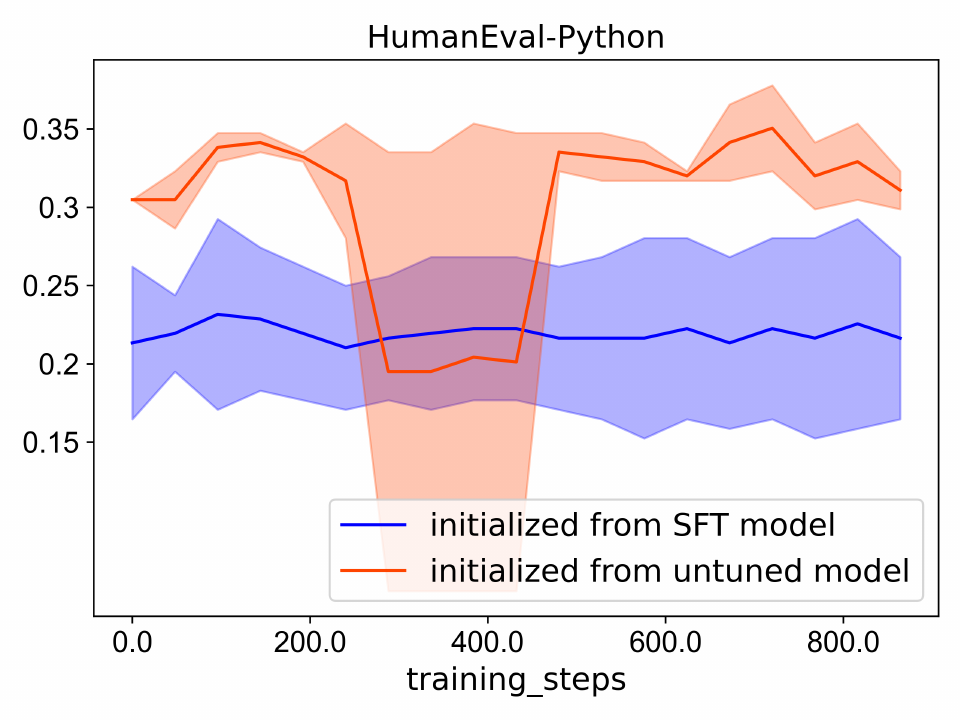} \hfill
  \includegraphics[width=0.48\linewidth]{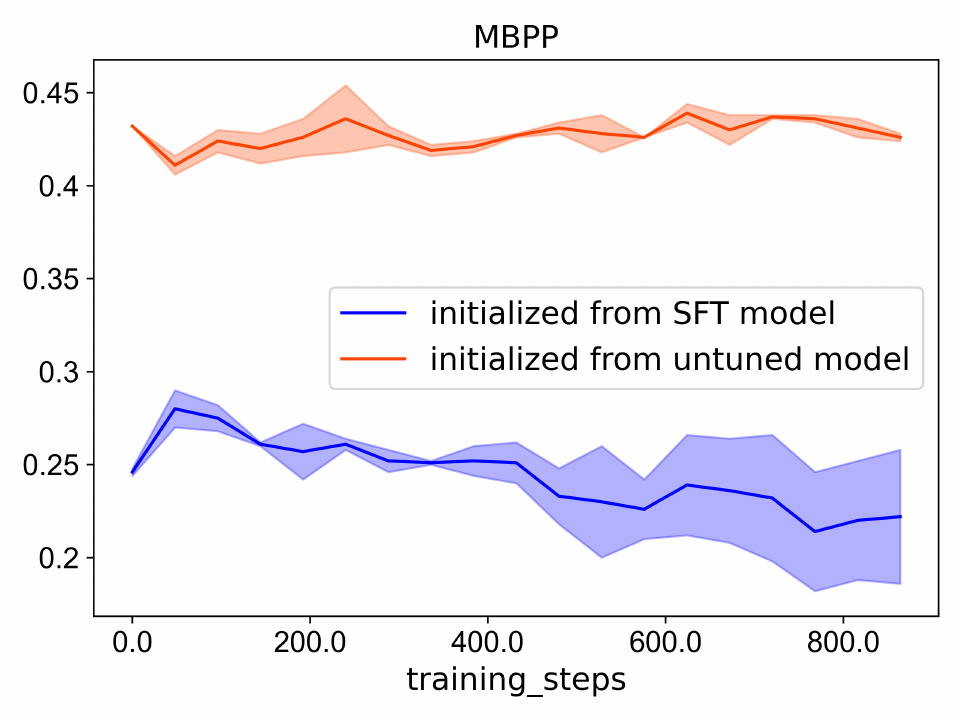}
  \quad
  \includegraphics[width=0.48\linewidth]{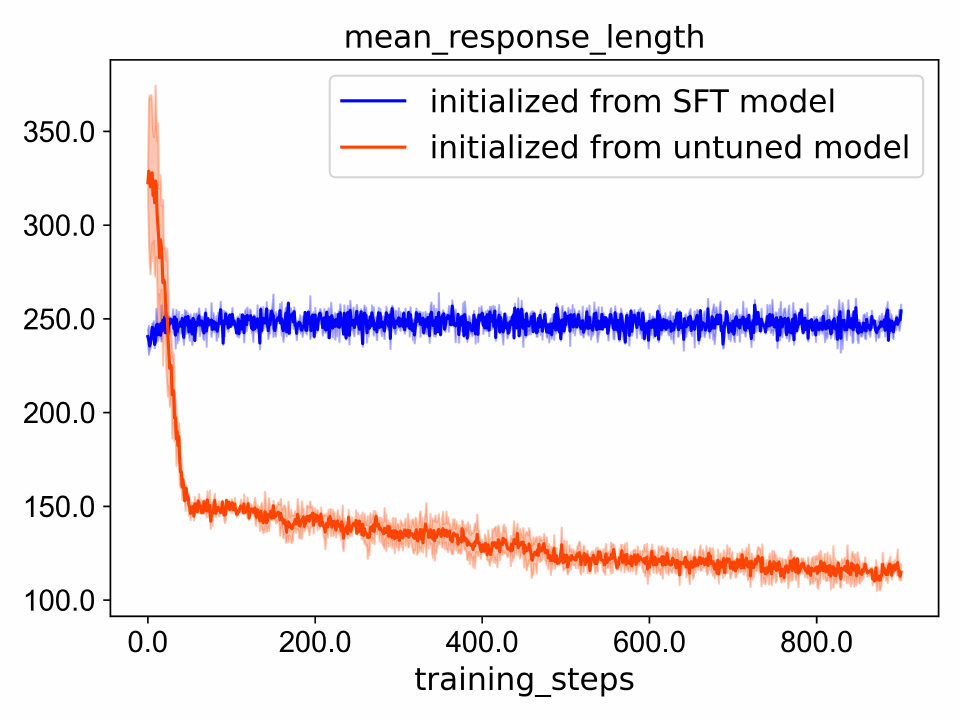} \hfill
  \includegraphics[width=0.48\linewidth]{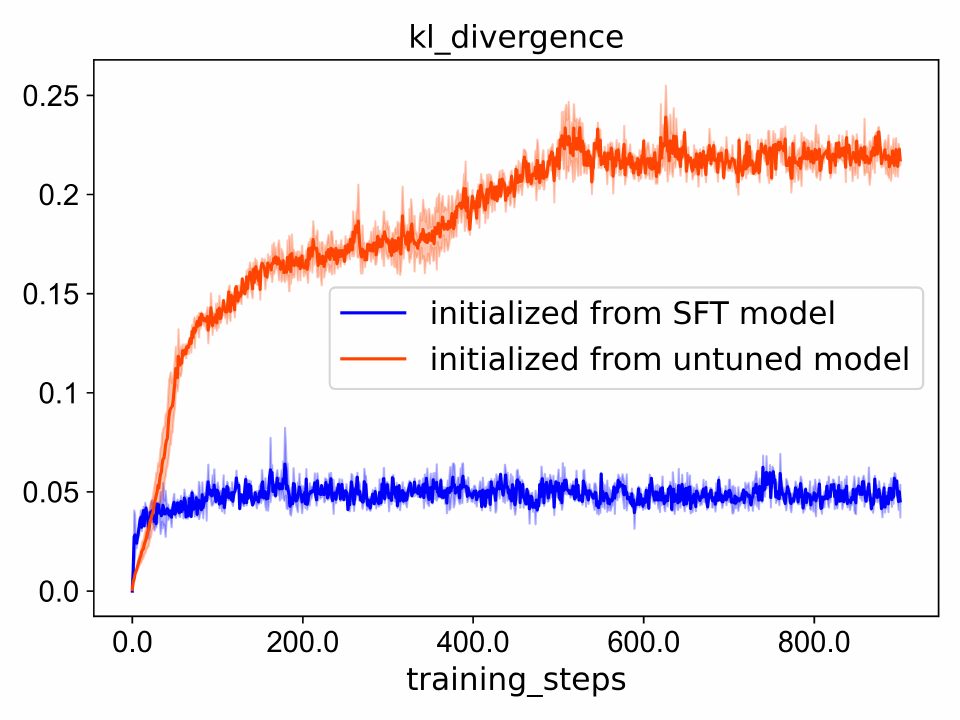}
  \quad
  \includegraphics[width=0.48\linewidth]{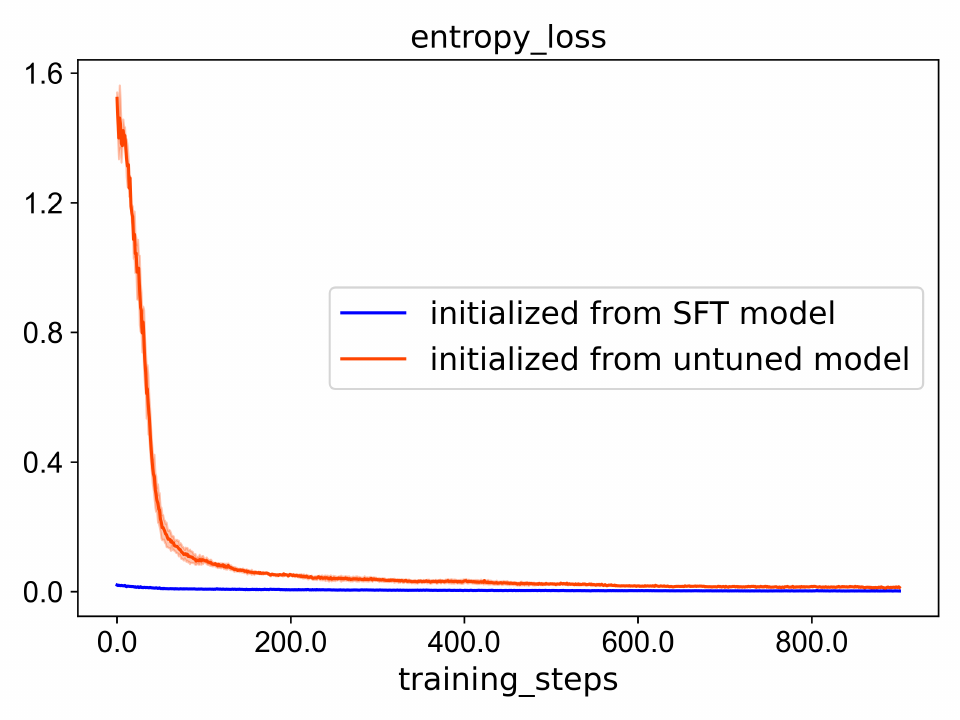} \hfill
  \includegraphics[width=0.48\linewidth]{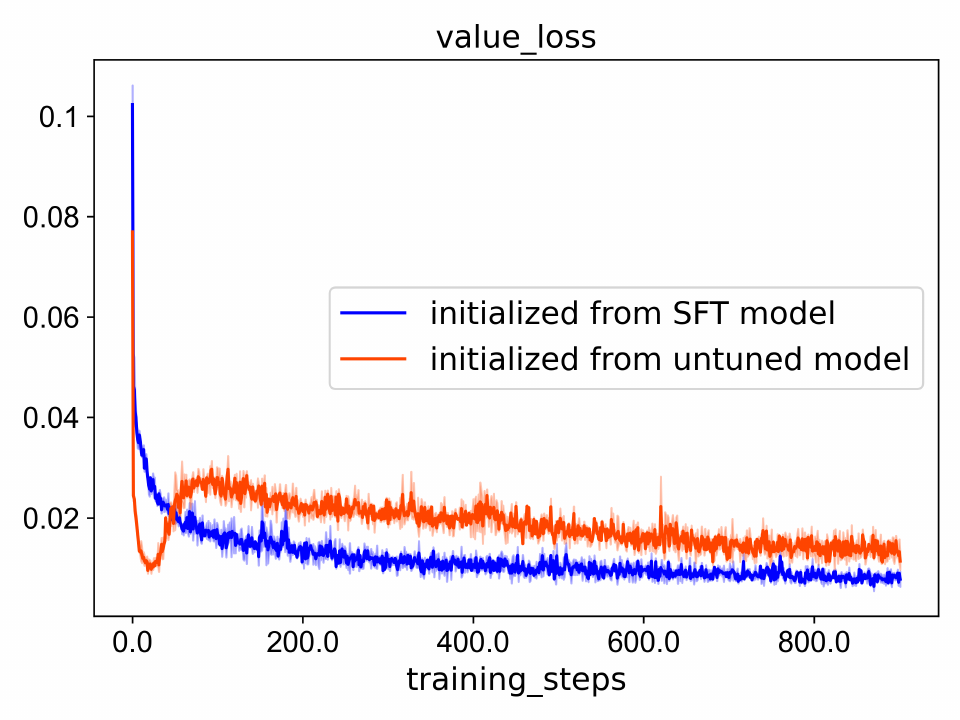}
  \captionof{figure} {Training details of reinforcement learning on Composite\_B with different initialized checkpoints. Initialized from the untuned model, the KL divergence, entropy loss and the value loss are higher, while the mean response length tends to be shorter.}
  \label{fig:diff_init_HardB_full}
\end{minipage}
\clearpage
\begin{minipage}{0.96\textwidth}
  \includegraphics[width=0.48\linewidth]{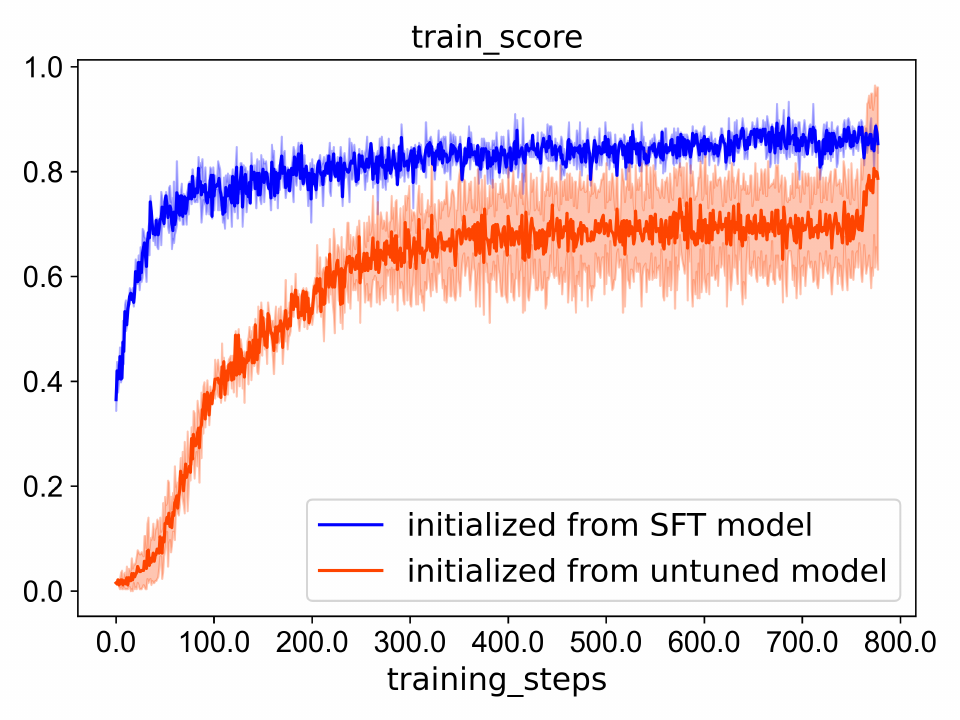} \hfill
  \includegraphics[width=0.48\linewidth]{RL_HardB_Atom_diff_init/test_score.pdf}
  \quad
  \includegraphics[width=0.48\linewidth]{RL_HardB_Atom_diff_init/HumanEval-Python.pdf} \hfill
  \includegraphics[width=0.48\linewidth]{RL_HardB_Atom_diff_init/MBPP.pdf}
  \quad
  \includegraphics[width=0.48\linewidth]{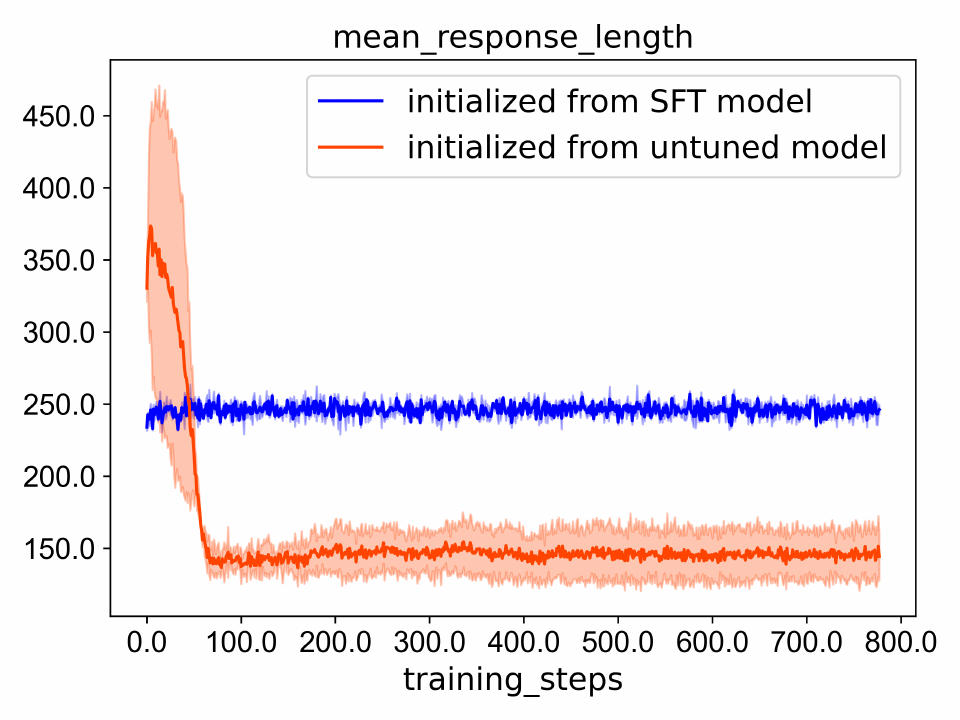} \hfill
  \includegraphics[width=0.48\linewidth]{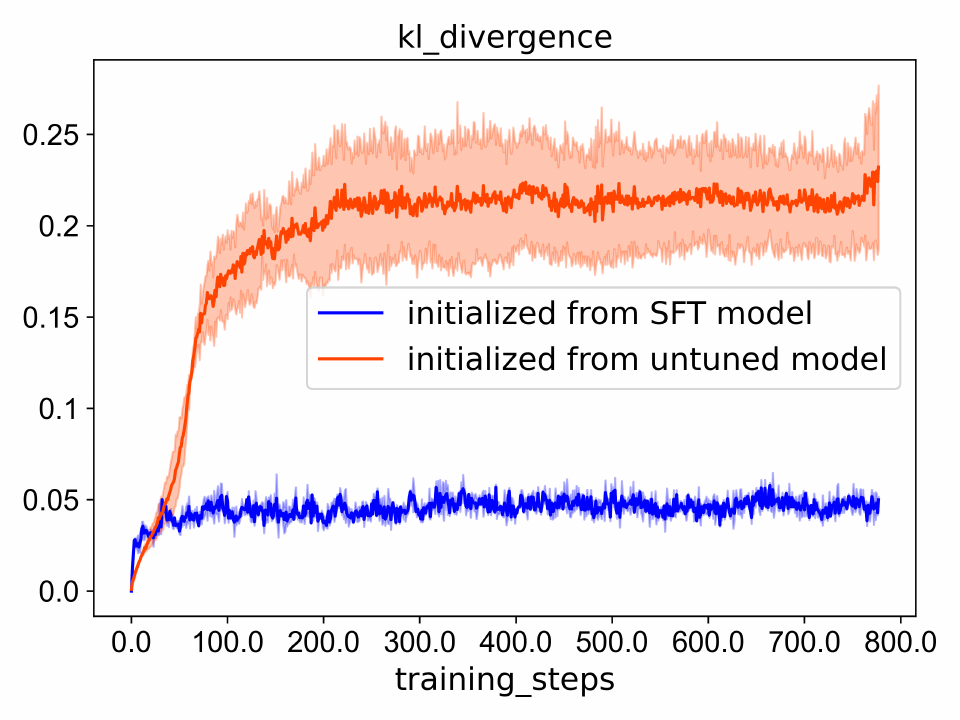}
  \quad
  \includegraphics[width=0.48\linewidth]{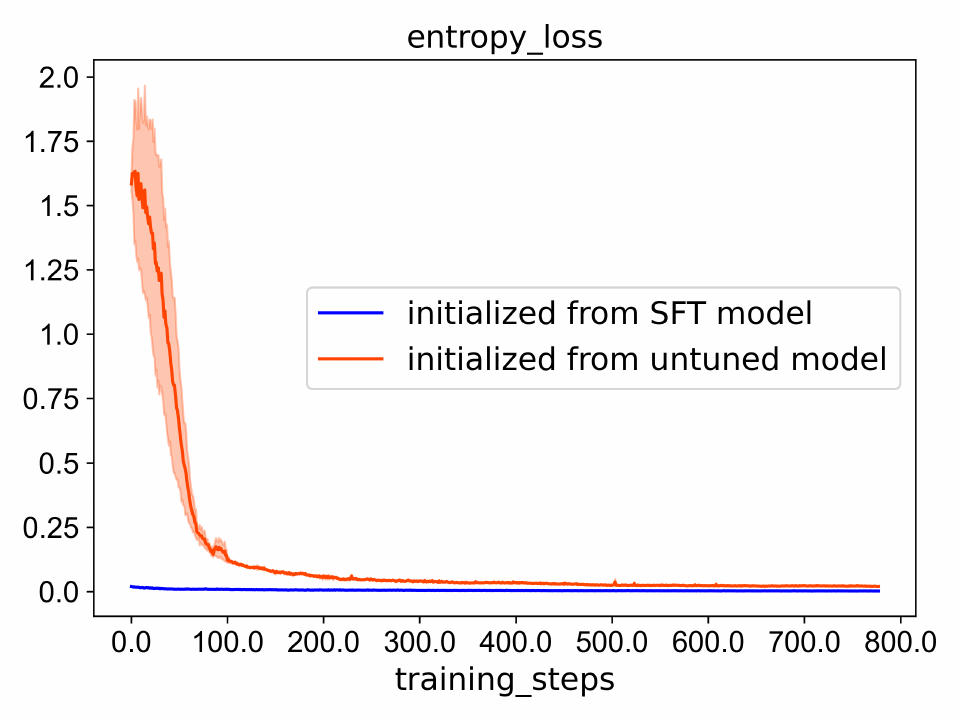} \hfill
  \includegraphics[width=0.48\linewidth]{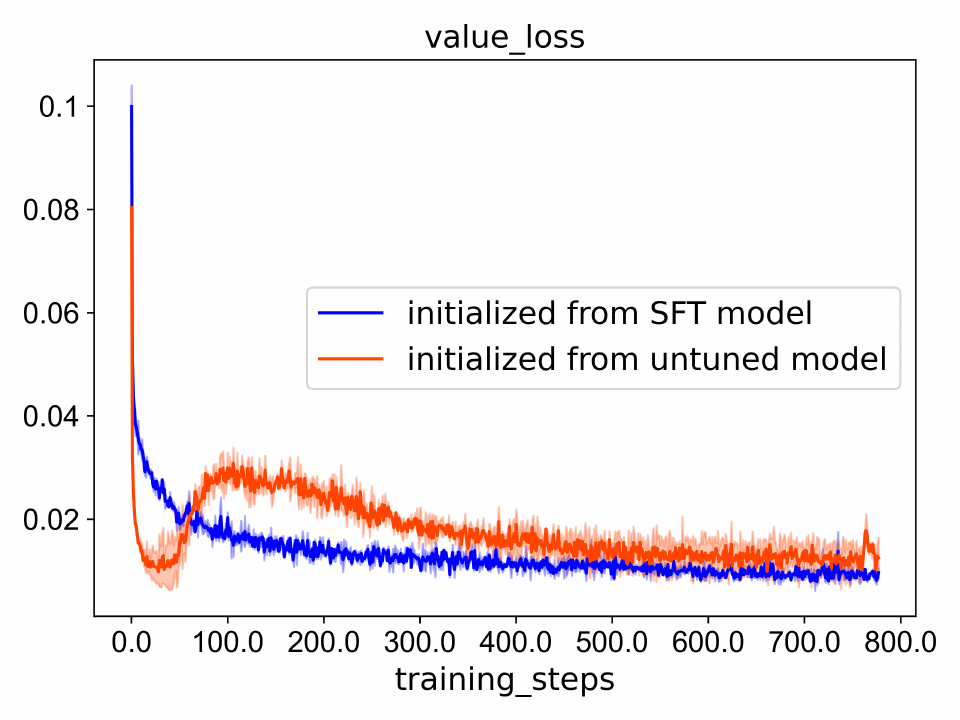}
  \captionof{figure} {Training details of reinforcement learning on Composite\_B and Atom\_base with different initialized checkpoints. Similar as ~\cref{fig:diff_init_HardB_full}, the KL divergence, entropy loss and the value loss are higher, and the mean response length is shorter, when initializing the RL training phase from the untuned model.}
  \label{fig:diff_init_HardB_Atom_full}
\end{minipage}

\clearpage
\section{Case Study for RL from Different Initializations}

In order to provide a detailed overview on how different initializations influences the performance of RL, we present a few representative cases in this section. Specifically, our RL is initialized from two different checkpoints, namely the SFT and the untuned ones, and is fine-tuned on the same training dataset, namely Composite\_B and Atom\_base. Therefore, we have two different RL models, to make it simple, we would like to refer to the two models as SFT-RL and untuned-RL in the following section.

\begin{figure*}
    \centering
    \includegraphics[width=1.0\textwidth]{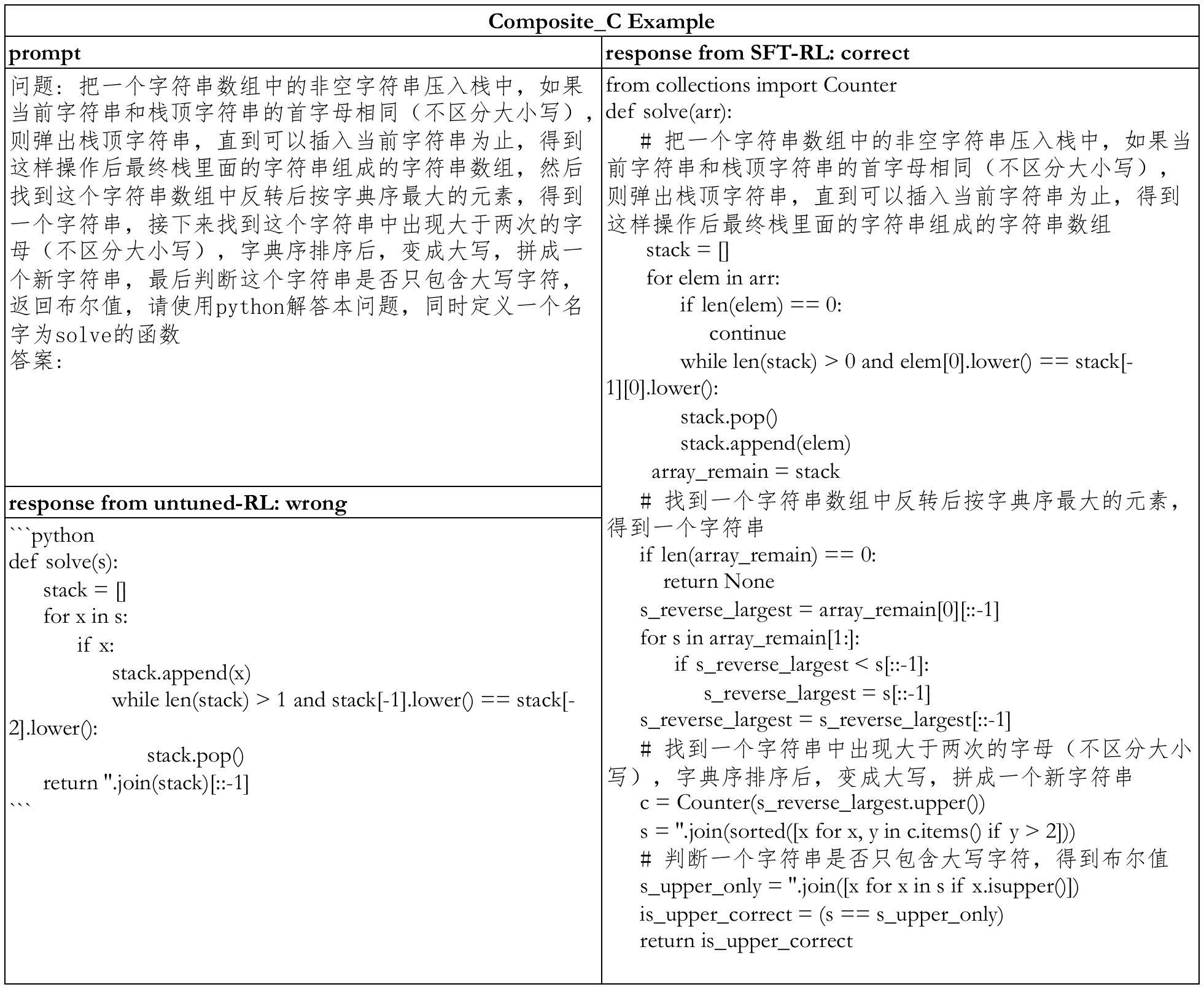}
    \strut\newline
    \centering
    \includegraphics[width=1.0\textwidth]{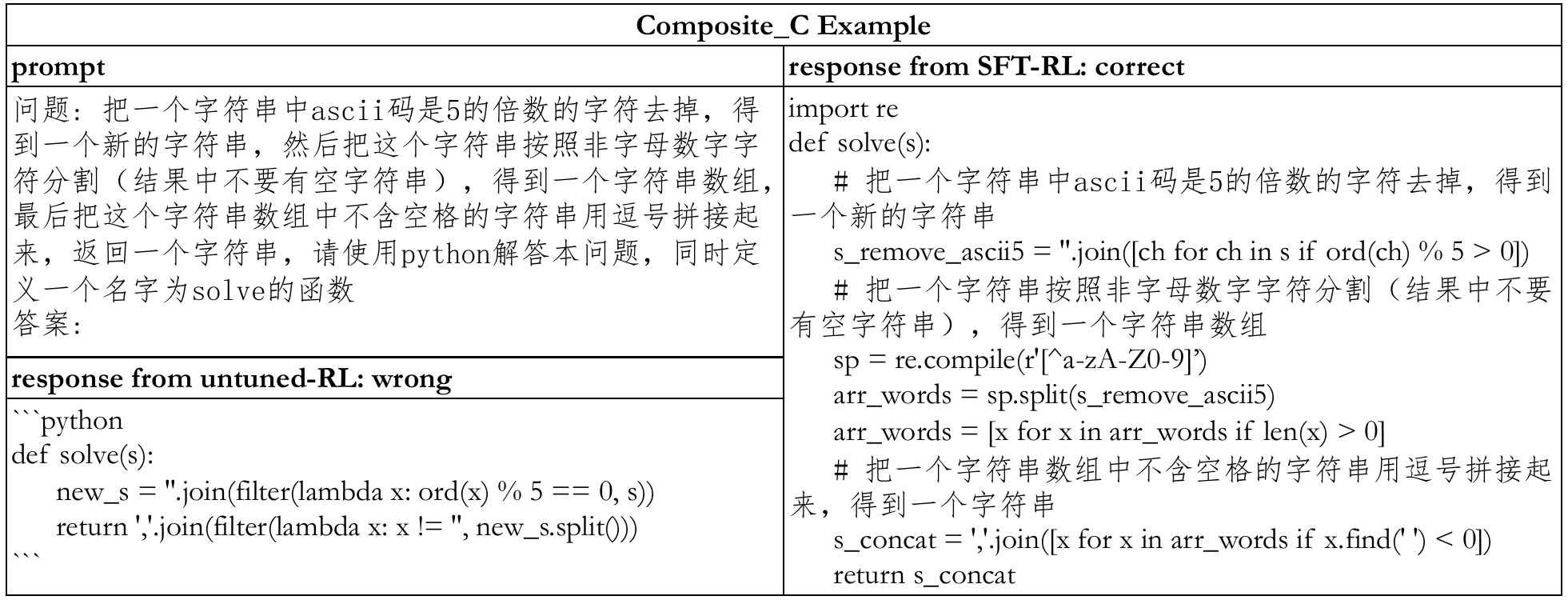}
    \caption{Cases from Composite\_C generated by SFT-RL and untuned-RL}\label{fig:case-hardc-1}
\end{figure*}

\begin{figure*}
    \strut\newline
    \centering
    \includegraphics[width=1.0\textwidth]{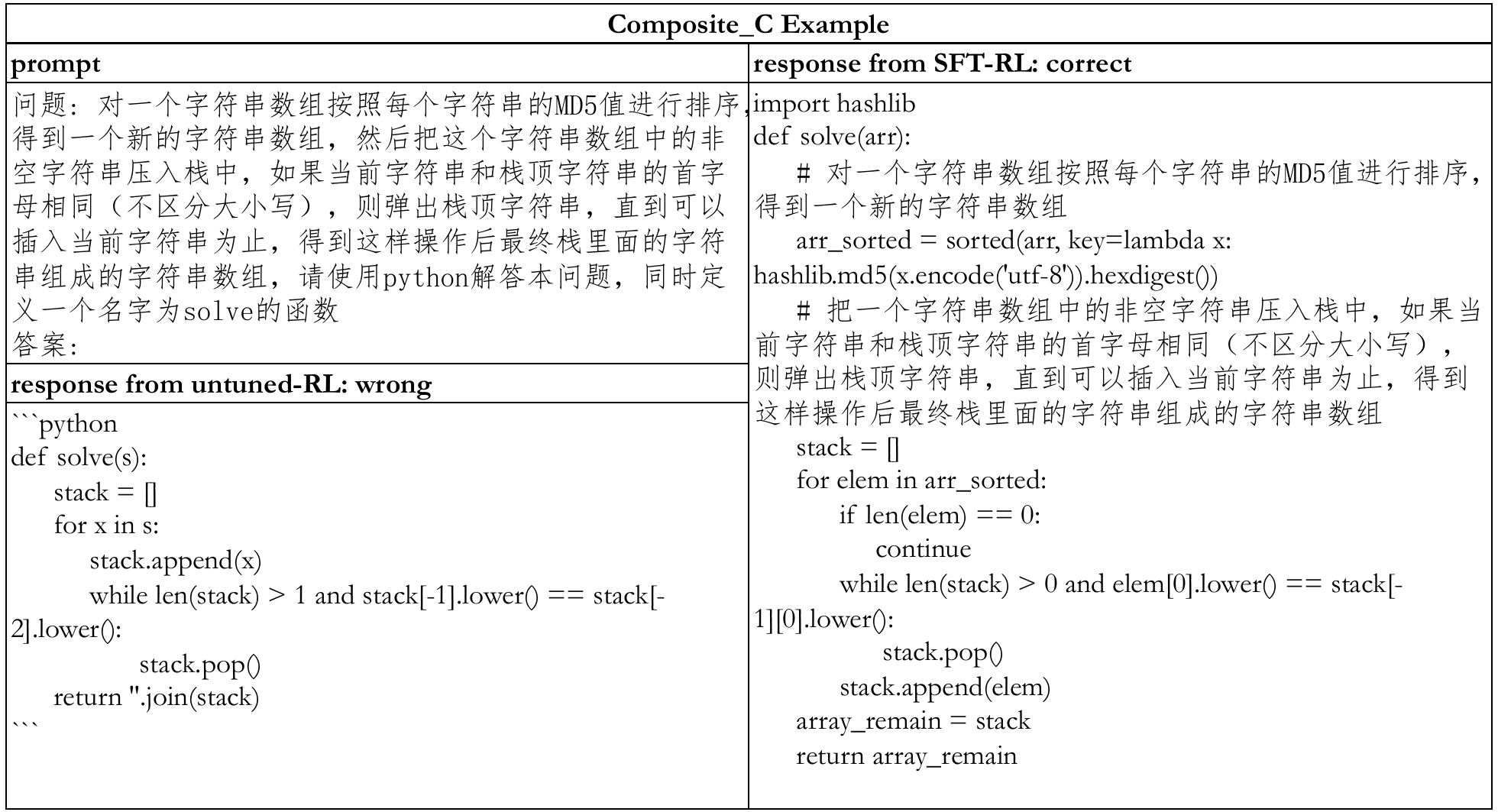}
    \strut\newline
    \centering
    \includegraphics[width=1.0\textwidth]{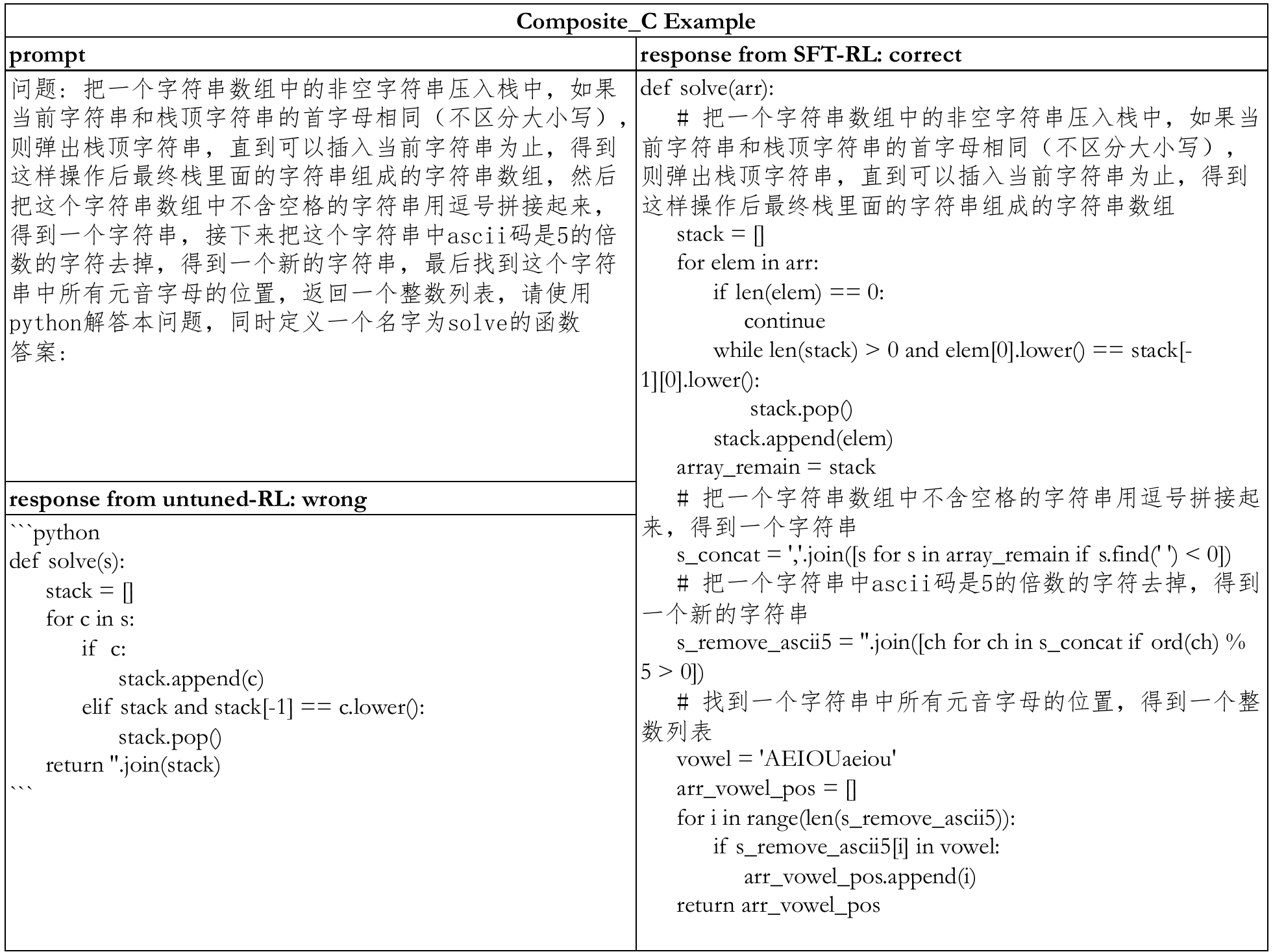}
    \caption{Cases from Composite\_C generated by SFT-RL and untuned-RL}\label{fig:case-hardc-2}
\end{figure*}

In \cref{fig:case-hardc-1} and \cref{fig:case-hardc-2}, cases from the Composite\_C dataset are shown. It is clear that the untuned-RL performs poorly than the SFT-RL and fails to solve all of the problems. Actually, to write the correct codes for problems contained in Composite\_C, the model needs to: (1) Understand the synthetic prompts and well split them into sub-prompts; (2) Generate the correct coding block for each sub-prompt; (3) Organize the corresponding coding blocks in the correct order and format. As displayed, the SFT-RL model can successfully accomplish the above three tasks. However, as for the untuned-RL model, it struggles to understand the presented synthetic prompts.

\begin{figure*}
    \centering
    \includegraphics[width=1.0\textwidth]{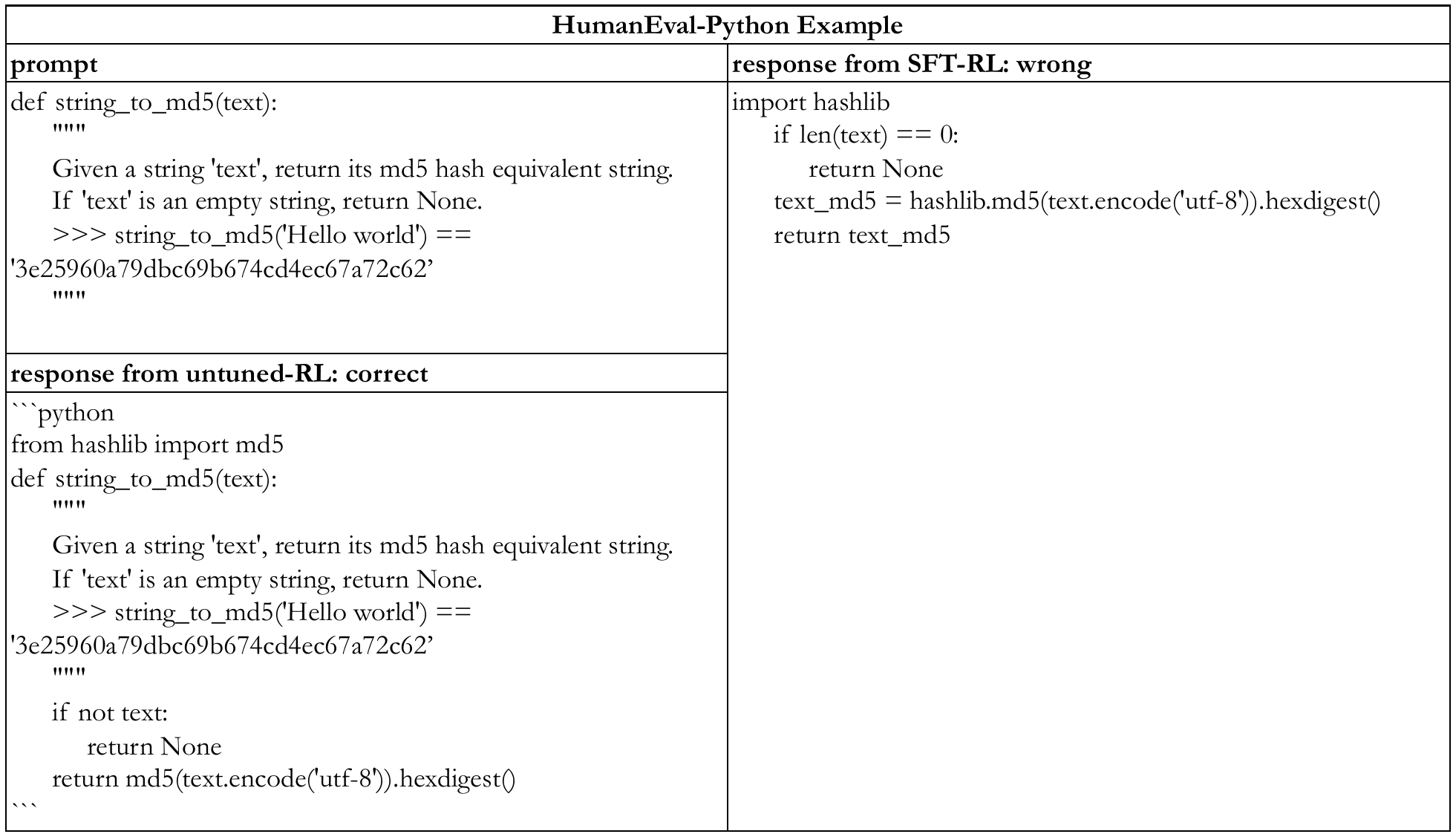}
    \strut\newline
    \centering
    \includegraphics[width=1.0\textwidth]{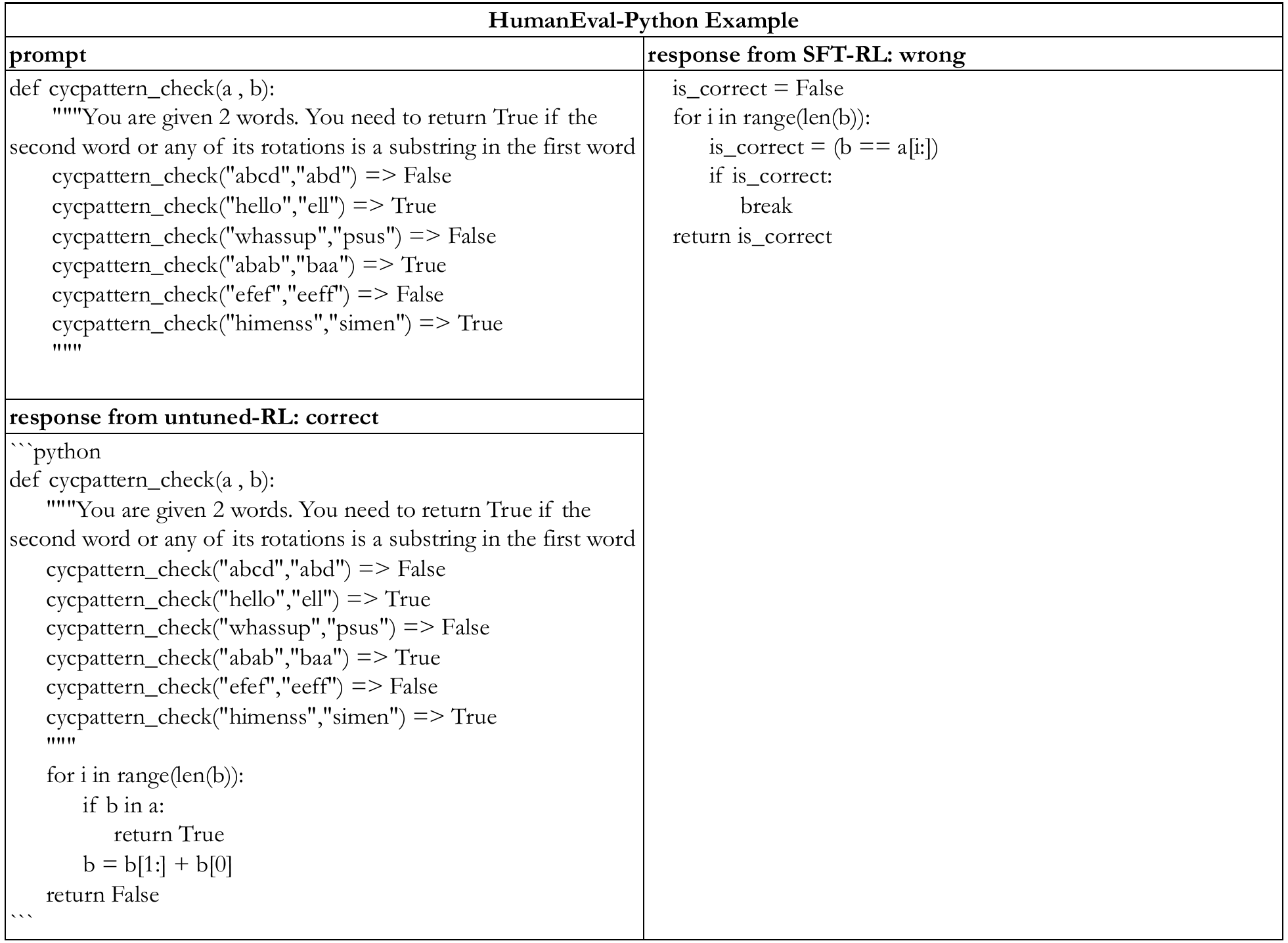}
    \caption{Cases from HumanEval generated by SFT-RL and untuned-RL}\label{fig:case-humaneval-1}
\end{figure*}

\begin{figure*}
    \centering
    \includegraphics[width=1.0\textwidth]{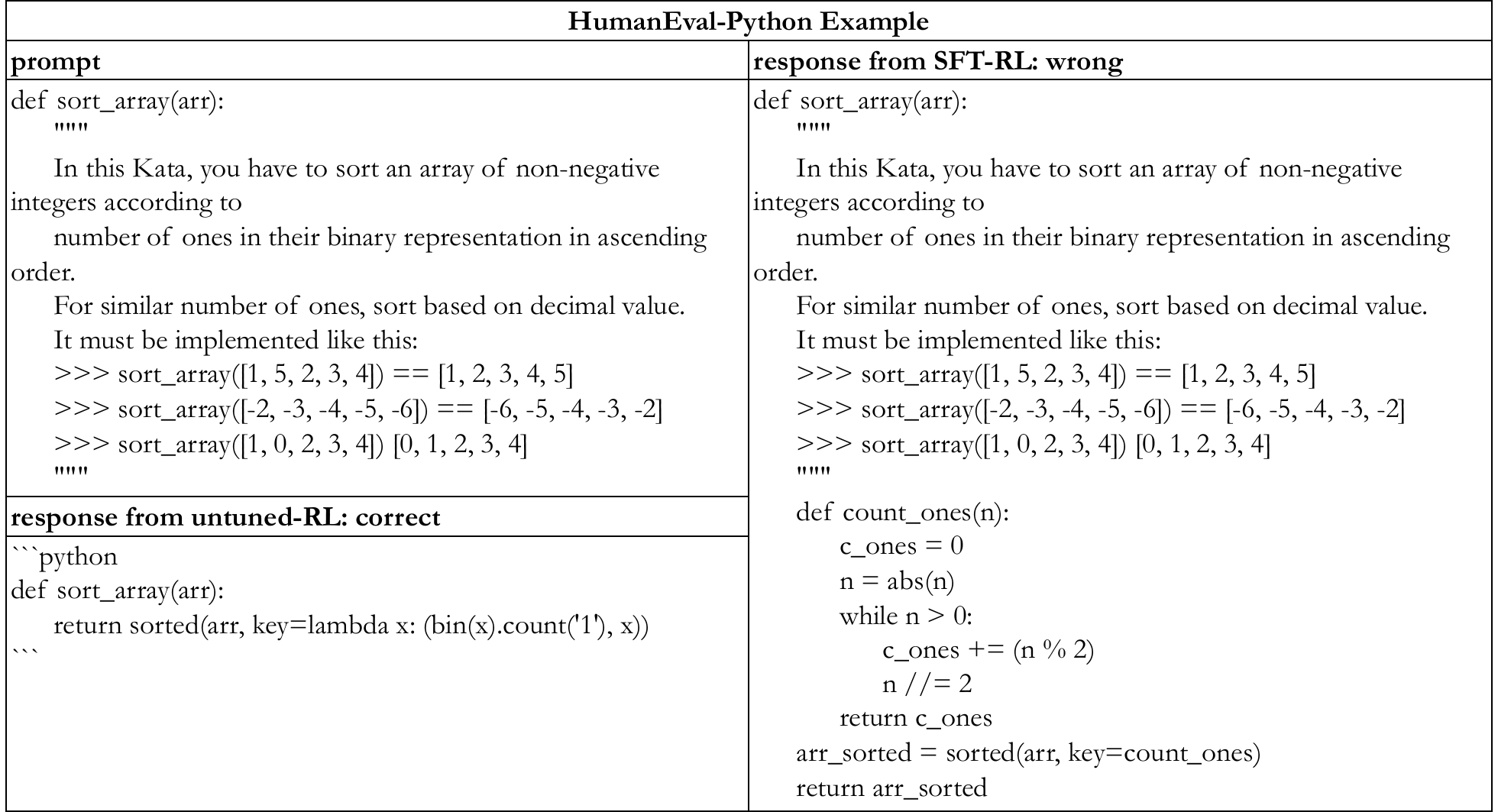}
    \strut\newline
    \centering
    \includegraphics[width=1.0\textwidth]{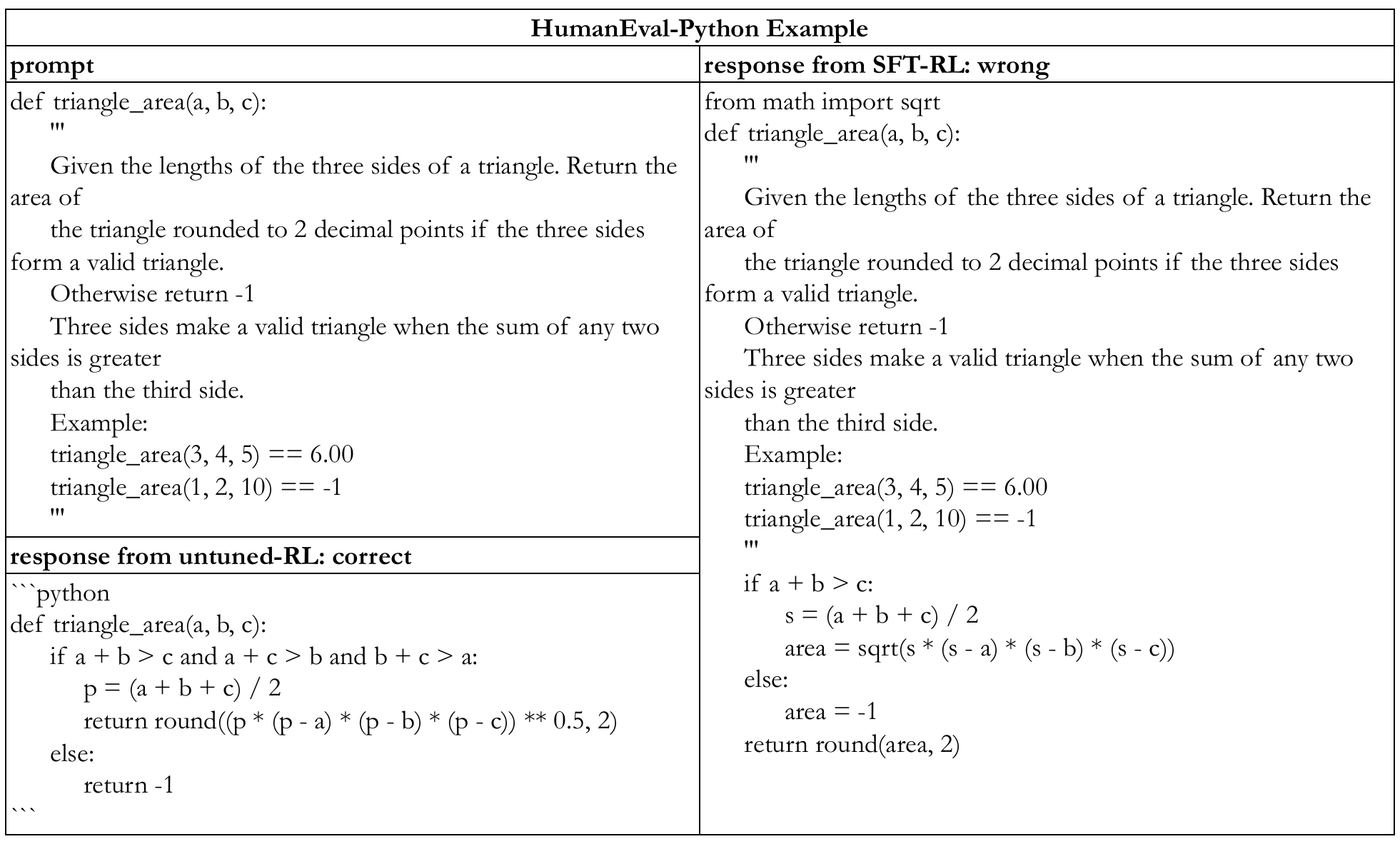}
    \caption{Cases from HumanEval generated by SFT-RL and untuned-RL}\label{fig:case-humaneval-2}
\end{figure*}

As pointed out in \cref{sec:rlfromuntuned}, although the SFT-RL model performs well on Composite\_C, if fails to generalize to HumanEval-Python and MBPP due to the overfitting issues. A few cases from the HumanEval-Python are presented in \cref{fig:case-humaneval-1} and \cref{fig:case-humaneval-2}. The prompts in HumanEval-Python are consisted of function signatures and the doc-strings. In order to generate the correct response, the model needs to copy the prompt firstly and then continue to write the code. However, the SFT-RL model tends to generate responses from the continuation point of a code-completion task more often, thus fail to produce an executable code block. The untuned-RL model performs better in HumanEval-Python, and it wraps the responses within a Python markdown block.

\begin{figure*}
    \centering
    \includegraphics[width=1.0\textwidth]{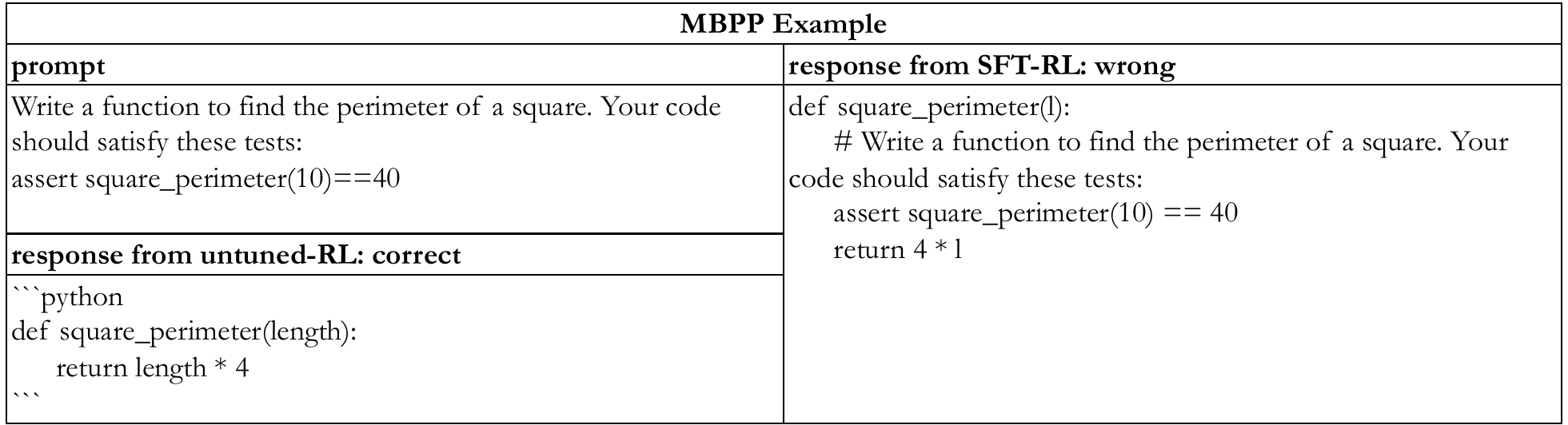}
    \vspace{0.2cm}
    \newline
    \centering
    \includegraphics[width=1.0\textwidth]{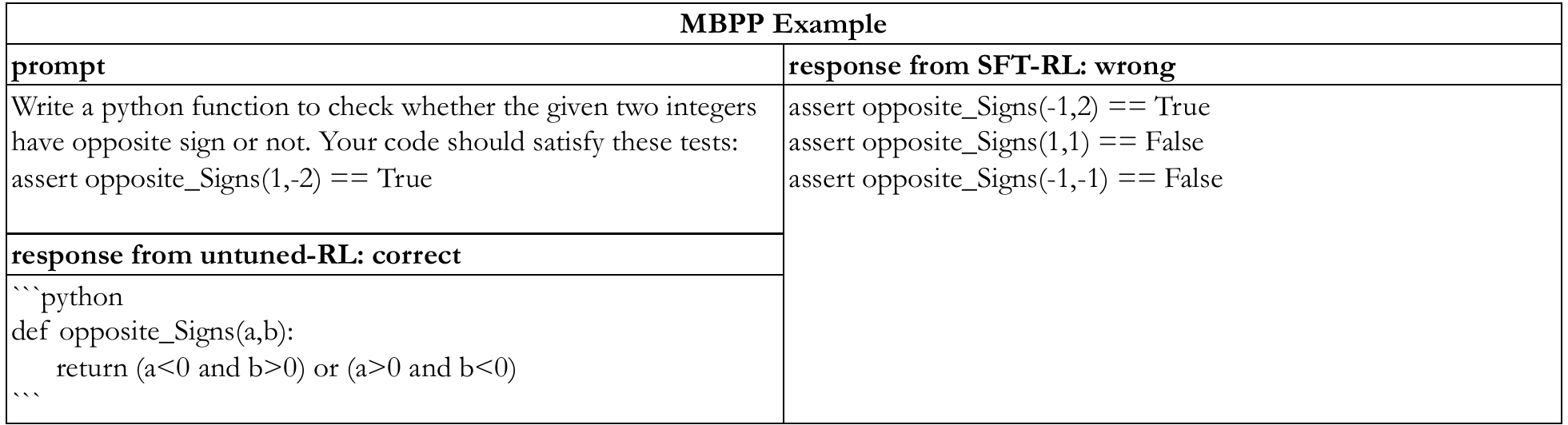}
    \vspace{0.2cm}
    \newline
    \centering
    \includegraphics[width=1.0\textwidth]{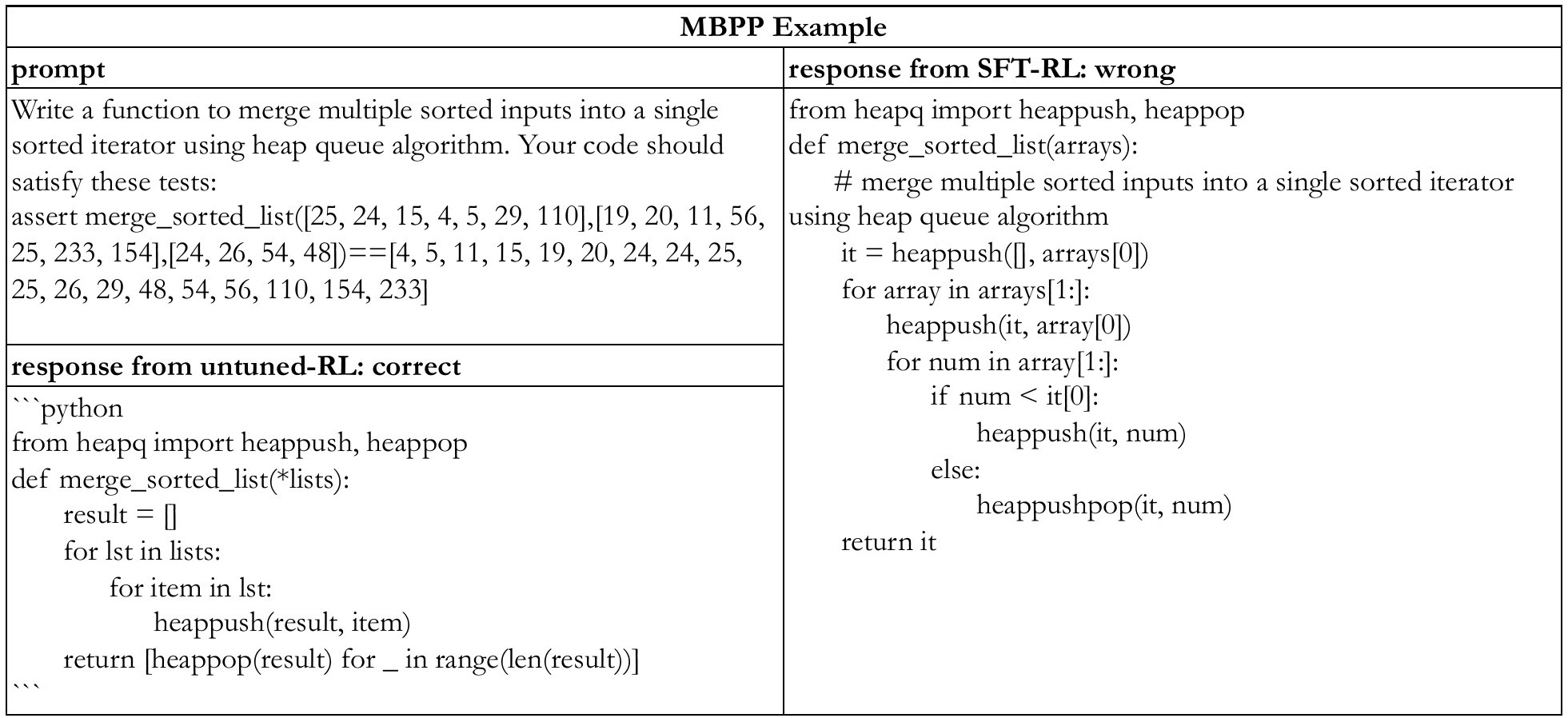}
    \vspace{0.2cm}
    \newline
    \centering
    \includegraphics[width=1.0\textwidth]{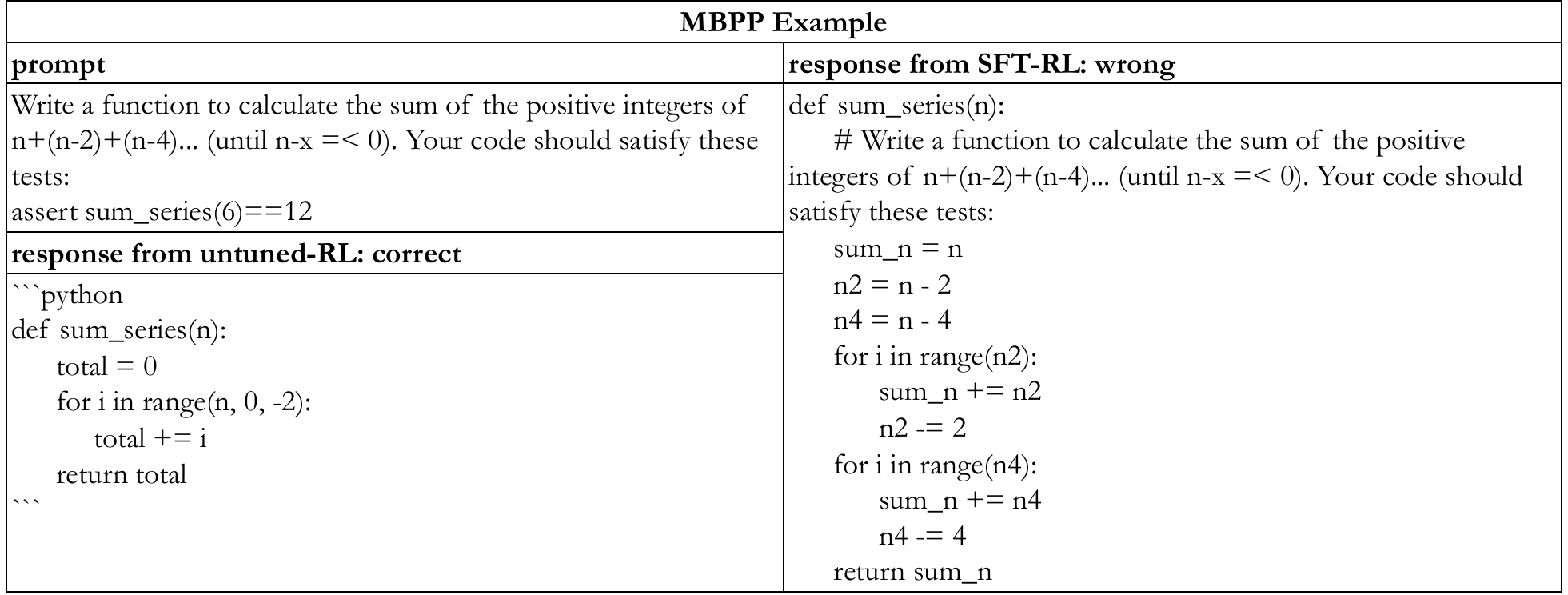}
    \caption{Cases from MBPP generated by SFT-RL and untuned-RL}\label{fig:case-mbpp}
\end{figure*}

Cases from the MBPP test dataset are shown in \cref{fig:case-mbpp}. Prompts in MBPP often contains a natural language instruction followed by a unit test case, and the model is required to generate codes matching the instructions while passing all the test cases. As seen, the untuned-RL model performs well on MBPP, similarly as in HumanEval-Python, it tends to wrap the generated responses using a Python markdown block. Moreover, the programming style of the untuned-RL model is straight forward and neat. However, the SFT-RL model fails to solve most problems in MBPP. In some cases, it is clearly influenced by the "assert" expression in the prompt and generates hallucination by inserting wrong assertion expressions in the code.

In the cases from Composite\_C, HumanEval-Python, and MBPP, we can observe that the untuned-RL model tends to generate much shorter responses compared to that of the SFT-RL model. We have included the "mean\_response\_length" as a metric in \cref{fig:diff_init_HardB_full} and \cref{fig:diff_init_HardB_Atom_full}, which calculates the mean length of the generated responses within a batch. Clearly, the untuned-RL model generates much shorter responses during training. Besides, the KL divergence of the untuned-RL model with respect to the initialized checkpoint is much higher than that of the SFT-RL, indicating more exploration has been achieved during training. Meanwhile, the entropy loss and the value loss of the SFT-RL model are much lower and more stable than that of the untuned-RL. 

\clearpage
\clearpage

\clearpage
\clearpage

\clearpage
\clearpage

\clearpage
\clearpage

\clearpage
\clearpage

\end{document}